\documentclass[12pt,aps,preprint,nofootinbib]{revtex4}

\usepackage{graphicx}
\usepackage{cancel}
\usepackage{amssymb}
\usepackage{textcomp}
\usepackage{amsmath}
\usepackage{bm}
\usepackage{times}
\usepackage{epsfig}
\usepackage{color}

\def\lsim{\mathrel{\raise.3ex\hbox{$<$\kern-.75em\lower1ex\hbox{$\sim$}}}}
\def\gsim{\mathrel{\raise.3ex\hbox{$>$\kern-.75em\lower1ex\hbox{$\sim$}}}}

\def\beq{\begin{equation}}
\def\eeq{\end{equation}}
\def\be{\begin{equation}}
\def\ee{\end{equation}}
\def\bea{\begin{eqnarray}}
\def\eea{\end{eqnarray}}

\def\etmiss{\cancel{E}_{T}}

\def\dm{\Delta M}
\def\mi{M_1}
\def\mii{M_2}
\def\miii{M_3}
\def\mg{M_3}

\def\mlsp{M_{LSP}}

\def\mnone{M_{\chi^0_1}}
\def\mntwo{M_{\chi^0_2}}

\def\mcone{M_{\chi^\pm_1}}

\def\mw{M_W}
\def\mz{M_Z}

\def\gev{\,{\rm GeV}}

\def\to{\rightarrow}

\begin{document}
\preprint{CERN-PH-TH/2010-056}
\preprint{~~MADPH-10-1556}
\preprint{~~IPMU09-0102}

\title{Nearly Degenerate Gauginos and Dark Matter at the LHC}

\author{Gian F.~Giudice$^{\bf a}$}
\email{gian.giudice@cern.ch}

\author{Tao Han$^{\bf b}$}
\email{than@hep.wisc.edu}

\author{Kai Wang$^{\bf c}$}
\email{kai.wang@ipmu.jp}

\author{Lian-Tao Wang$^{\bf d}$}
\email{lianwang@princeton.edu}

\affiliation{$^{\bf a}$  CERN, Theory Division, CH-1211 Geneva 23, Switzerland\\ 
$^{\bf b}$  Department of Physics, University of Wisconsin, Madison, WI 53706, USA\\
$^{\bf c}$  Institute for the Physics and Mathematics of the Universe, University of Tokyo, 
Kashiwa, Chiba 277-8568, Japan\\
$^{\bf d}$  Department of Physics, Princeton University, Princeton, NJ
08540, USA}

\begin{abstract}

Motivated by dark-matter considerations in supersymmetric theories, 
we investigate in a fairly model-independent way the detection 
at the LHC of nearly degenerate gauginos with mass differences between a few GeV and about $30$ GeV.
Due to the degeneracy of gaugino states, the conventional leptonic signals are likely lost.
We first consider the leading signal from gluino production and decay.
 We find that it is quite conceivable to reach a large statistical significance
 for the multi-jet plus missing energy signal  with an integrated luminosity about 50  pb$^{-1}$
(50  fb$^{-1}$) for a gluino mass of 500 GeV (1 TeV). If gluinos are not too heavy, less than about 1.5 TeV, 
 this channel can typically probe gaugino masses up to about 100 GeV below the gluino mass.
 We then study the Drell-Yan type of gaugino pair production in association with
 a hard QCD jet,  for gaugino masses in the range of  $100$--$150$ GeV.
 The signal observation may be statistically feasible with about 10  fb$^{-1}$,
 but systematically challenging due to the lack of distinctive features for the
 signal distributions.  By exploiting gaugino  pair production 
 through weak boson fusion, signals of large missing energy plus two forward-backward
 jets may be observable at a $4$--$6\sigma$ level above the large SM backgrounds 
 with an integrated luminosity of $100$--$300$ fb$^{-1}$. 
 Finally, we point out that  searching for additional isolated soft muons
 in the range $p_T \sim 3$--$10$ GeV in the data samples discussed above 
 may help to enrich the signal and to control the systematics.
Significant efforts are made to explore the connection between 
the signal kinematics and the relevant masses for the gluino and gauginos, 
to probe the mass scales of the superpartners, in particular the LSP dark matter.  

\end{abstract}

\maketitle

\section{Introduction}
 
If supersymmetry (SUSY) is realized in nature, and the SUSY partners of the standard model (SM)
particles are present at the weak scale, then new colored supersymmetric particles will be copiously produced 
at the LHC via the SU(3)$_{\rm color}$ strong interaction. 
However, the definitive confirmation of supersymmetry will require the discovery of the supersymmetric partners of the electroweak SM particles as well.The identification of the electroweak sector of the supersymmetric theory and the measurement of its parameters is especially important because it is believed that the dark matter particle, the ``Lightest Supersymmetric Particle'' 
(LSP),  resides in this sector.
On the other hand, the direct production of electroweak supersymmetric particles at the LHC suffers from relatively small rates, while the indirect production in decay chains is rather model-dependent, rendering the 
missing particle identification and its property determination challenging.

A further complication is that, whenever the soft SUSY breaking mass parameters are larger than weak boson mass $M_W$, some of the charginos and neutralinos become nearly degenerate in mass, making their identification at the LHC more problematic. For instance, when the LSP is mostly wino, as in models with anomaly mediation~\cite{amsb}, the mass difference between the lightest chargino and neutralino is, in the limit of large $\mu$,
\beq
\mcone - \mnone \simeq \frac{\mw^4 \sin^2 2\beta \tan^2\theta_W}{(\mi -\mii )\mu^2} +\frac{\alpha \mw}{2(1+\cos \theta_W)}.
\label{masdifw}
\eeq
For large $\tan\beta$ the tree-level contribution in Eq.~(\ref{masdifw}) is suppressed and the leading effect comes only at ${\cal O} (\mii M_W^4/\mu^4)$. Larger mass splittings can be obtained by introducing higher-dimensional operators suppressed by an intermediate scale~\cite{Berkooz:2008gw}. 

In the opposite case in which the gaugino masses are larger than $\mu$, the LSP is mostly higgsino, and two neutralinos and one chargino are approximately degenerate with mass differences
\beq
\mntwo -\mnone \simeq 2\left( \mcone - \mnone \right) \simeq \left( \frac{1}{\mii}+\frac{\tan^2\theta_W}{\mi}\right) \mw^2,
\label{masdifh} 
\eeq
where for simplicity we have taken the limit of large $\tan\beta$. The one-loop corrections to Eq.~(\ref{masdifh}) are larger than in the case of the wino, because the leading effect comes from top-stop loops.

Another possibility is that $\mi$ is accidentally very close to either $\mii$ or $\mu$, making the bino nearly degenerate in mass with other states. This case may not seem generic in the allowed parameter space of soft SUSY masses, 
but it is actually motivated by dark matter considerations. Indeed, the annihilation rates of higgsinos and winos in the early universe are too fast to make these particles good cold dark matter particles as thermal relics, unless their masses are larger than one TeV and thus beyond the region favored by naturalness considerations of the weak-scale SUSY. 
On the other hand, the annihilation rate of binos is typically insufficient to account for a dark-matter thermal relic, and requires an enhancement from a coannihilation channel. The mixed cases of bino-wino or bino-higgsino are therefore particularly important, due to the fact that they correctly reproduce the required thermal relic abundance. Mixed neutralinos with masses in the range between 100 and 300 GeV are acceptable dark matter candidates if the relative mass splittings are less than about $10\% -20 \%,$ depending on the specific case \cite{Arkani-Hamed:2006mb}.  

For this reason it is quite important to investigate the collider search strategies for scenarios in which some neutralinos and charginos are degenerate in mass at the level of 10--20 GeV or less. 
Quite often the final state LSP, which is the dark matter particle and escapes detection, comes from the decay of the nearly degenerate ``Next-to-Lightest Supersymmetric Particle'' (NLSP), and thus the accompanying decay products (SM leptons and quarks) are rather soft, typically not passing the detector acceptance, and thus becoming unobservable. 
Even if the colored supersymmetric particles, such as the gluino, are light, the clean leptonic modes may be lost. 
Therefore, it is necessary to re-evaluate 
the experimental signatures of this scenario and check the observability at hadron colliders.

In this article, we explore the signatures of nearly degenerate electroweak gauginos at the LHC and, for concreteness, we mostly focus on the case of mixed bino-wino. The mass difference between NLSP and LSP is typically larger than about 1~GeV,  
and the NLSP thus decays promptly with in the detector.\footnote{For smaller mass differences, one will be led to the signatures of long-lived charginos, as in the case of pure wino LSP \cite{amsb_pheno}. We will not pursue such an analysis here.} 
We consider two classes of signatures:
\bea
&& {\rm class\ I:}\quad {\rm jets}+\etmiss ,\\
&& {\rm class\ II:}\quad  {\rm jets}+\etmiss +{\rm soft\ charged\ leptons}\ \ell^{\pm} .
\eea
The hard jets and large missing energy ($\etmiss$) serve as event triggers. 
The jet multiplicity depends on the underlying production and decay channel under consideration,
and the $\etmiss$ is not only from the LSP, but also directly from the NLSP which may not produce detectable 
decay products  or a displaced vertex.
In the second class of signal, the soft charged leptons resulting from the NLSP decay,
$\chi_1^\pm ,  \chi_2^0  \to \chi_1^0 \ell^\pm$'s, 
may not pass the triggering requirements, but can be searched for  with off-line analyses
of those events. 
In addition to expanding the discovery reach for the gauginos, this class of observables can 
be particularly important in measuring the properties of the LSP and thus discriminating between the mixed bino-Higgsino and bino-wino cases. 

In order to focus on the most model-independent features of the signal, 
we consider the conservative limit in which there are 
no light squarks or sleptons to enhance the supersymmetric production rates. This situation is explicitly realized in models with heavy scalars~\cite{pev} or in Split Supersymmetry~\cite{S-SUSY}.
Although squarks and sleptons are assumed to be 
out of reach of the LHC, gluinos may still  be accessible. We thus first consider the leading channel
of production and decay
\beq
pp \to \tilde{g} \tilde{g} \to q q \chi^0_{i},\  q q^{'} \chi^\pm_{j}.
\label{gluinos}
\eeq
The signature in the above process  would typically lead to four jets from light quarks plus large missing energy.
Given the small mass difference of the order of $\dm \lsim 10-25$ GeV,
the charged leptons from the NLSP decay may be too soft to lead to striking signatures.
The detection of such soft leptons, however, would provide more convincing evidence for the scenario 
of degenerate gauginos under consideration.
The gauginos from heavy gluino decays are also boosted which makes the lepton transverse momentum
($p_T^{\ell}$) depend not only on the mass difference but also on the gluino mass itself. 
We explore the feasibility of observation for this channel at the LHC in Sec.~II.

With or without the contribution from gluino production in  the process  of Eq.~(\ref{gluinos}), the electroweak 
gaugino pairs can be produced by the standard electroweak processes 
\bea
&& pp \to \chi_1^\pm \chi_1^\mp \ X \to \ell^\pm \ell^\mp + \etmiss , \nonumber \\
&& pp \to \chi_1^\pm \chi_2^0 \ X \to \ell^\pm \ell^\mp \ell^\pm + \etmiss , \label{ewpp}
\eea
often leading to di-lepton and tri-lepton signals for SUSY.
However, for nearly degenerate gauginos, these clear signals are lost
because the charged leptons are too soft.
We are forced to consider these pair production
processes in association with a hard QCD jet to trigger on. We will study this mono-jet
plus large missing energy signal, as well as possible soft leptons, in Sec.~III.

Alternatively, we can consider gaugino pair production via the weak boson fusion (WBF) mechanism 
\beq
qq' \to qq'  \chi_1^\pm \chi_1^\mp,  \ \  \chi_1^\pm \chi_2^0.
\eeq
The characteristic  feature of these processes is the energetic accompanying
jets in the forward-backward region with transverse momenta of the order of  $M_W/2$.
This motivates the ``forward-jet tagging",  along with the requirement of large $\etmiss$.
Another important feature of the WBF processes is the absence of color exchange 
between the final state quarks, which leads to a suppression of gluon emission 
in the central region between the two tagging jets. 
We can thus enhance the signal to background ratio by central jet vetoing.  While the WBF processes may not be the primary discovery channels for degenerate gauginos, they will be very important to probe the gaugino properties. 
The production rates for the WBF processes are very different for bino, wino and higgsino, or mixed scenarios.  Therefore, together with signals from the other channels, even the observation or non-observation of degenerate gauginos in these channels provides valuable information. 
We will study this signal in Sec.~IV. 

The numerical studies of this paper are primarily performed for LHC with $E_{\rm CM} = 14$ TeV. 
The main effect of running at a lower c.m. energy is the sharp reduction of the production rate.  We will
compare the total cross sections at two c.m.~energies of 14 TeV and 7 TeV, and include relevant estimates and comments of the difference in signal reach. Our conclusions are given in Sec.~V.

\section{Gluino Pair Production}

\begin{figure}[tb]
\includegraphics[scale=2,width=8.1cm]{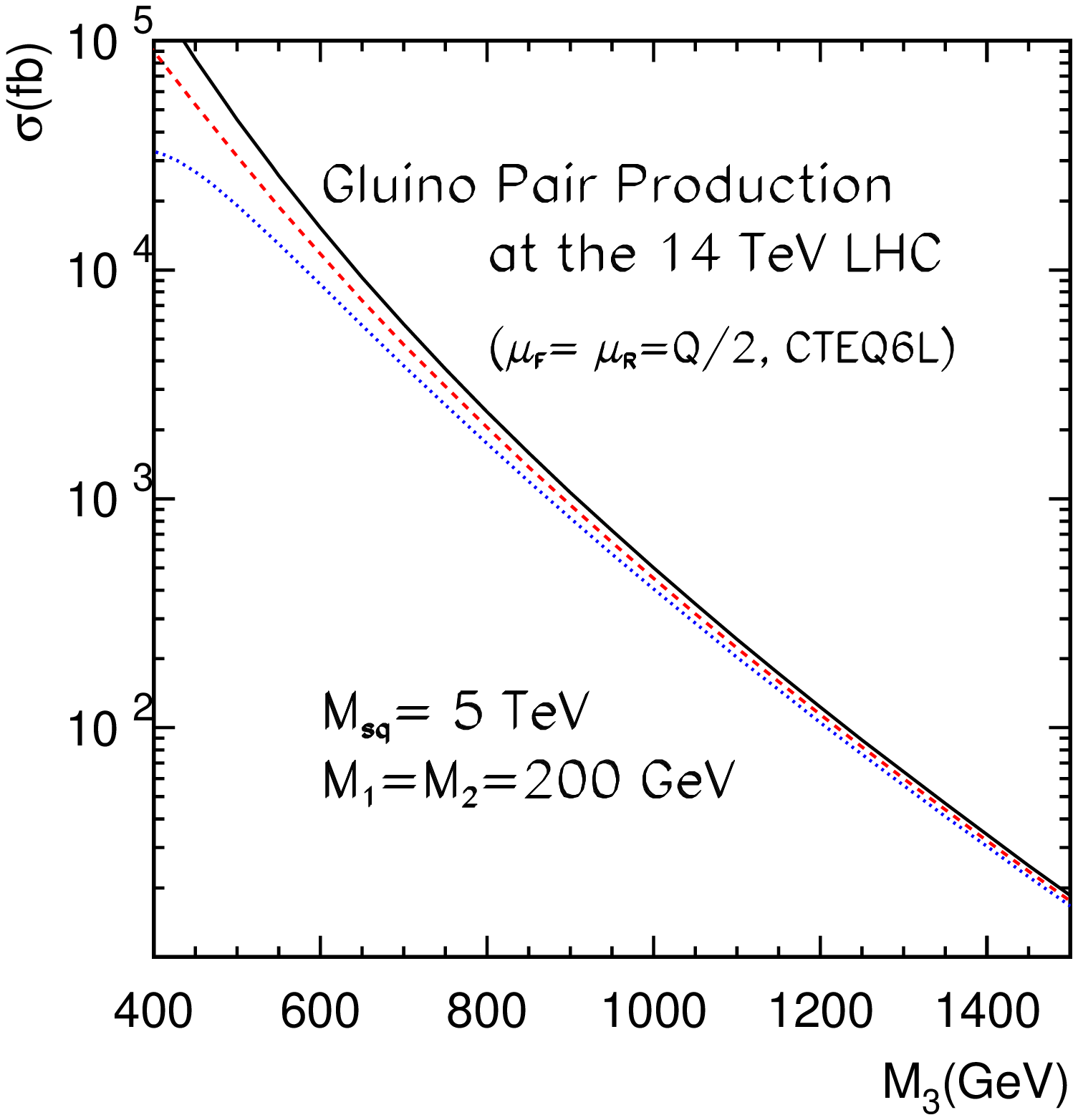}
\includegraphics[scale=2,width=8.1cm]{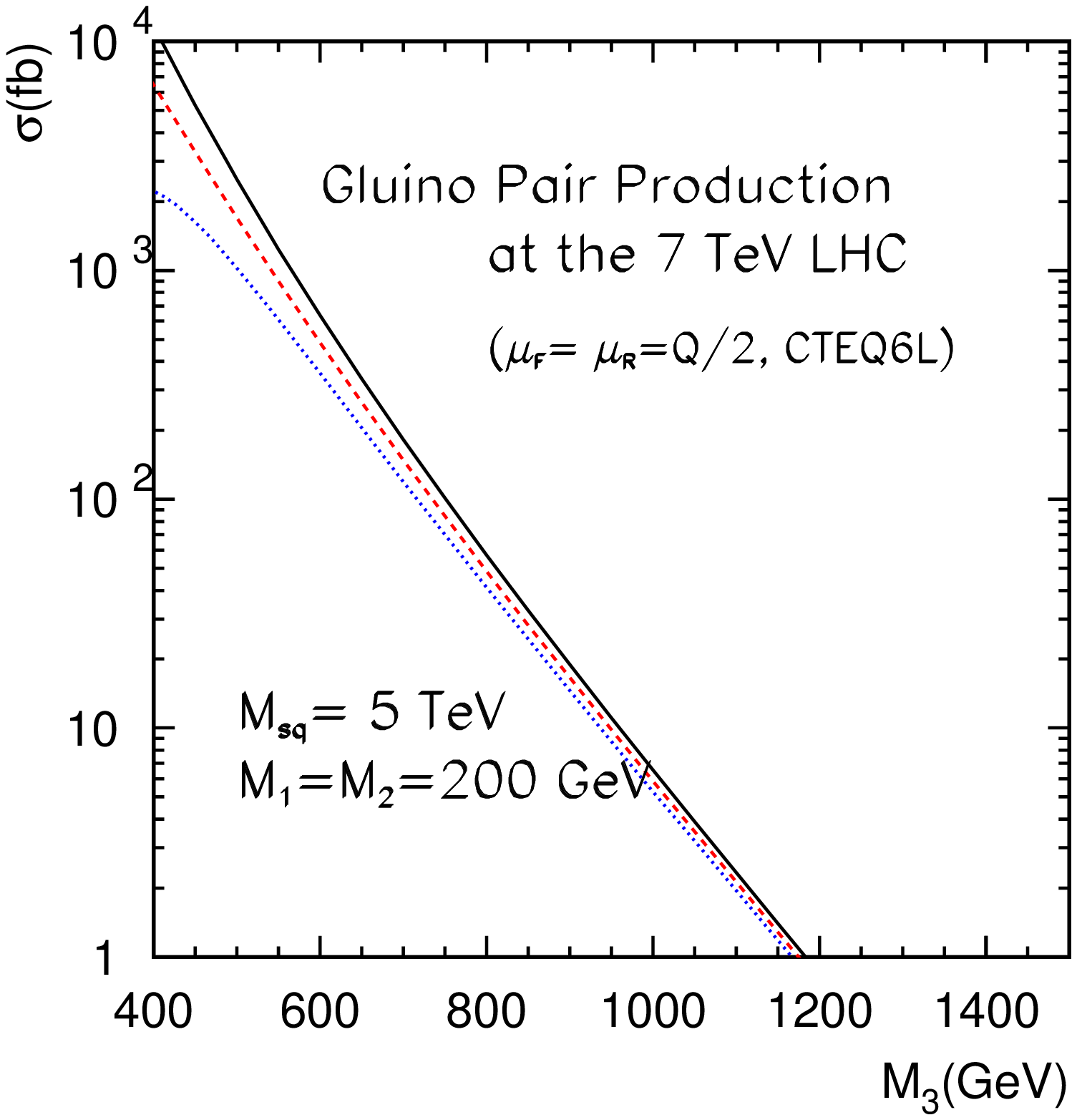}
\caption{Total cross section of gluino pair production versus the gluino mass $M_3$
for very heavy squarks ($M_{\tilde f}=5$ TeV) at the LHC for (a) 14 TeV  and (b) 7 TeV. 
The solid curves correspond to the case without kinematical cuts imposed. The dashed curves 
show the case with the missing energy cut of Eq.~(\ref{ETmiss}), and the dotted curves include
in addition the basic selection cuts in Eq.~(\ref{basic}).}
\label{fig:ggtot}
\end{figure}

Gluino pair production is usually considered to be one of the most important channels in SUSY searches at hadron colliders  due to the large production cross section
from QCD and, in particular, the large gluon luminosity at higher energies. 
The total cross section for gluino pair production is shown
as a function of the gluino mass  by the solid curve in Fig.~\ref{fig:ggtot}  at the LHC for the c.m.~energies
of (a) 14 TeV and (b) 7 TeV, with a very heavy squark mass. We see that the production cross section at the lower
energy of 7 TeV is decreased by more than an order of magnitude at a low gluino mass and becomes even
more suppressed at a higher mass.

We have used parton distribution functions (PDF) CTEQ6L \cite{Pumplin:2002vw} in our simulation. 
We use the SUSY MadGraph \cite{Cho:2006sx} and SDecay \cite{sdecay} for SUSY study
and MadGraph/MadEvent \cite{madgraph} for SM simulation.
The factorization scale and the renormalization scale in $\alpha_s$ are set to be equal, and 
taken to be $\mg$ for the signal, which is the gluino mass at the weak scale,
and to be $\sqrt{\hat{s}}/2$ for the background processes.
Since our main goal is to propose search strategies based on kinematical considerations, both
the signal and SM background calculations are only at tree level without including Next-Leading-Order
QCD corrections, and we have not included parton shower and matching. 
The quantitative result may be modified when taking into account those effects \cite{pt-isr}, while we expect the qualitative
features and conclusions to remain valid. 

\subsection{Model  Parameters}

To further demonstrate  general features of the gluino pair production signal 
in the degenerate gaugino limit,  we focus on the $\tilde B - \tilde W$ mixing case,
characterized by $M_{1}\simeq M_{2}$.
We choose two sets of parameters for the soft-SUSY breaking masses of the electroweak gauginos
\begin{eqnarray}
\label{mixi}
{\rm  Set\ I:}
&& M_1 = 120~\text{GeV},\quad M_2 =120~{\rm GeV} - 150~{\rm GeV},
\\
\label{mixii}
{\rm  Set\ II:}
&& M_1 = 200~\text{GeV},\quad M_2 =200~{\rm GeV} - 250~{\rm GeV},
\end{eqnarray}
with additional common parameters 
\be
\mu = 1~\text{TeV},\quad \tan{\beta} = 5, \quad A_{i} \simeq 0~\text{GeV},
\ee
and gluino and squark masses
\be
M_3=500~{\rm GeV} - 1500~{\rm GeV},\quad M_{\tilde{f}} = 5~\text{TeV}. 
\ee
The motivation for the parameter choices is as follows. 
By setting $\mu$ as large as 1 TeV, the Higgsino states $\chi^{\pm}_{2}$, $\chi^{0}_{3}$, $\chi^{0}_{4}$
are all heavy and gluino decaying into Higgsino states will thus be kinematically suppressed or forbidden. 
To simplify the discussion and to ensure squark decoupling, we assume large squark masses and $A\simeq 0$~GeV.

\subsection{Gaugino Decays}

Gluinos decay through virtual squarks into quarks and gauginos
\bea
\tilde{g}\rightarrow q\tilde{q}^{*} \rightarrow q q \chi^0_{1,2}\ {\rm or}\ q q^{'} \chi^\pm_1.
\label{eq:gg}
\eea
%
\begin{figure}[tb]
\includegraphics[scale=1.5,width=9.5cm]{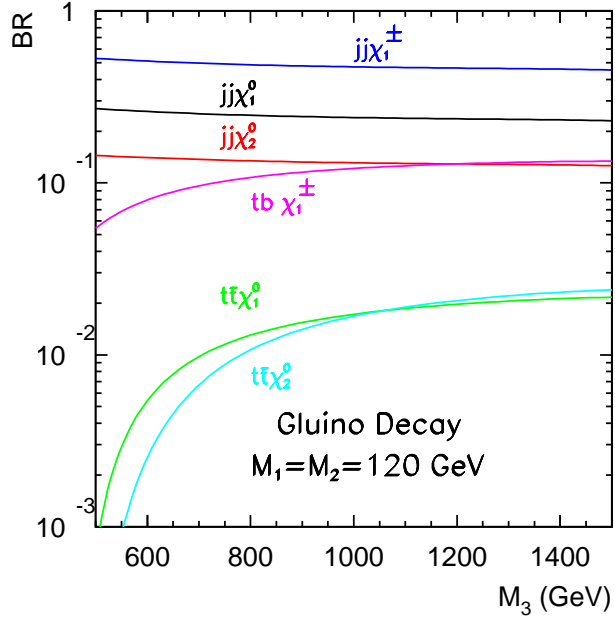}
\caption{Gluino decay branching fraction versus its mass for  $M_{1}=M_{2}=120$~GeV. 
A light-quark jet including $b$ is denoted by $j$. The channels involving a top are separately shown.
}
\label{fig:gluinodecay}
\end{figure}
Figure \ref{fig:gluinodecay} shows the 
gluino decay branching fractions versus its mass for $M_{1}=M_{2}=120$~GeV. 
A light-quark jet including $b$ is denoted by $j$ and more than $80\%$ of the BR goes to them. 
The channels involving a top quark are separately shown and the  phase space suppression is evident
for a lower $M_{3}$.  
However, since the partial width is proportional to $m_{\tilde{f}}^{-4}$, the decay branching fraction into the 3$^{rd}$ generation quarks can be significantly enhanced in scenarios which the masses of third generation squarks are somewhat smaller than those of the first two. Such a scenario leads to very different and interesting collider signals, featuring multiple lepton and multiple $b$ final 
states \cite{Baer:1990sc,Hisano:2002xq,Hisano:2003qu,Toharia:2005gm,Mercadante:2005vx,Gambino:2005eh,Baer:2007ya,Acharya:2009gb}. Here, we will focus on the more basic and more challenging scenario of gluino dominantly decaying into light quark jets.

\begin{figure}[tb]
\includegraphics[scale=1.5,width=8.1cm]{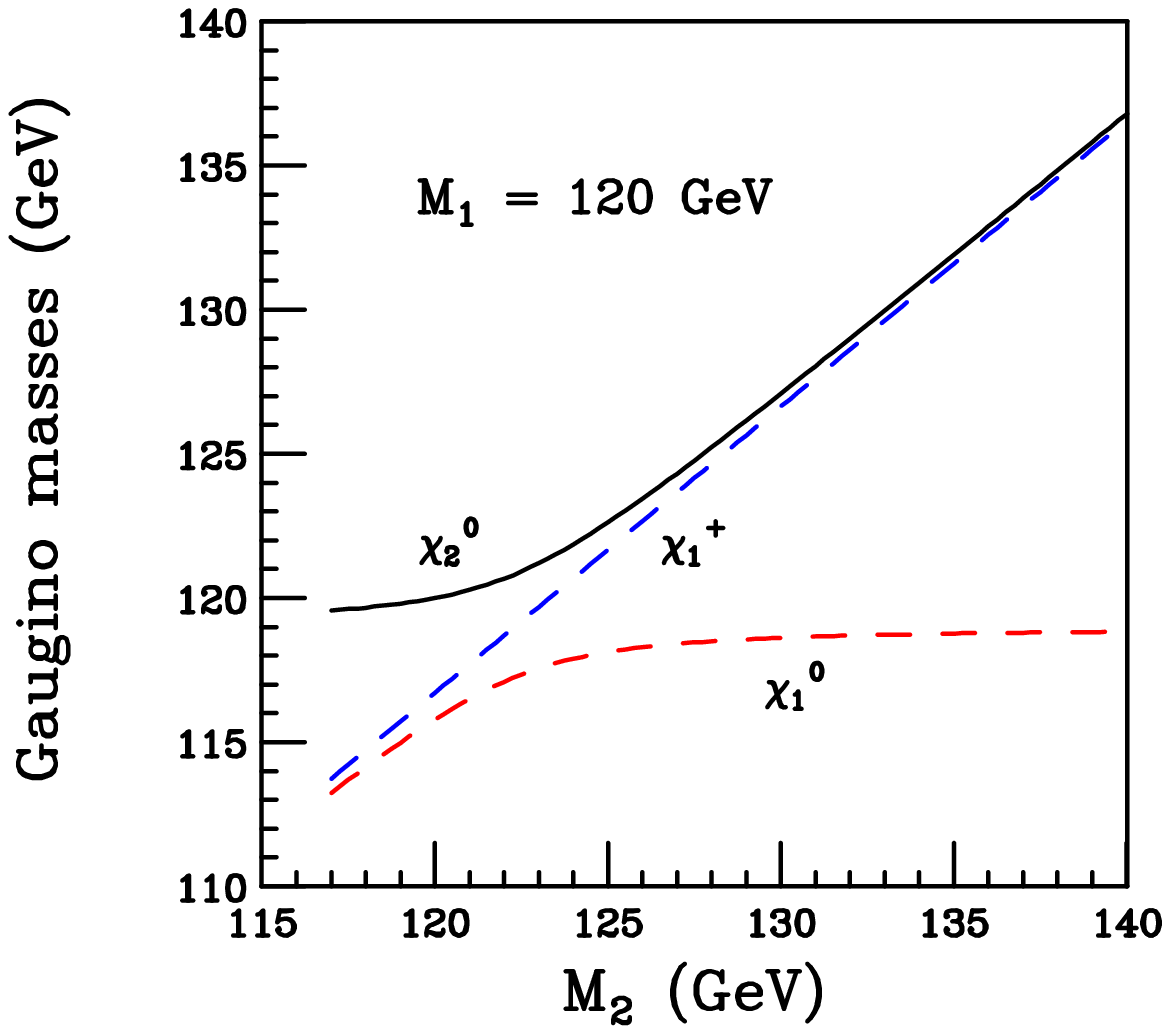}
\includegraphics[scale=1.5,width=8.26cm]{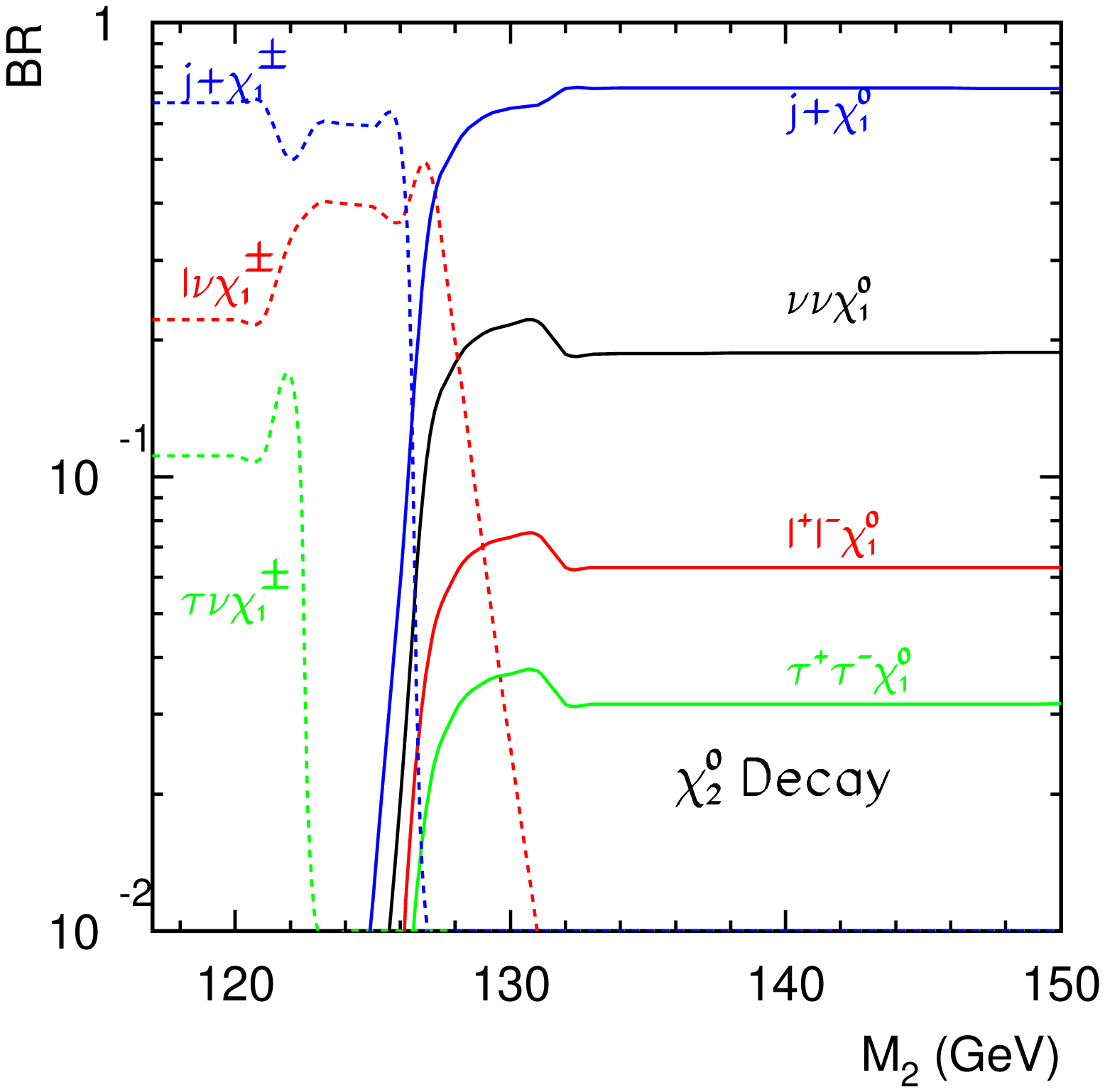}
\caption{(a) Lower-lying gaugino masses  and (b) $\chi^0_2$ 
decay branching fractions versus $M_{2}$ with $M_{1}=120$ GeV, 
with a light-quark jet denoted by $j$ (including $b$) and $\ell^\pm =e^\pm, \mu^\pm$. 
}
\label{fig:n2decay}
\end{figure}

The decay branching ratios of the electroweak  gauginos are governed by their mass difference. 
In Fig.~\ref{fig:n2decay}(a), we illustrate the lower lying gaugino masses versus $M_{2}$
for $M_{1}=120$ GeV.
The mass splittings between the gaugino states for $M_{2} > M_{1}$ are approximately given  by
\beq	
\dm \simeq \mcone -\mnone \simeq \mntwo - \mnone \simeq \mii - \mi -
\frac{\mz^2 \cos 2\theta_W\sin 2\beta}{\mu}  .
\eeq
Since the sfermions are set to decouple, $\chi^0_2$ and $\chi^\pm_1$ 
decay via virtual $W^{*}/Z^{*}$ as 
\bea
&& \chi^{0}_{2}\to \chi^{\pm}_{1} {W^{\mp}}^{*}\to\chi^{\pm}_{1}\ell^{\mp}\nu,\  \chi^{\pm}_{1} j j' ;\quad 
\chi^{0}_{2}\to\chi^{0}_{1} Z^{*} \to \ell^{+}\ell^{-}\chi^{0}_{1}, \ jj \chi^{0}_{1}, \\
&& \chi^{\pm}_{1} \to \chi^{0}_{1} W^{*} \to\ell^{\pm} \nu_{\ell} \chi^{0}_{1},\  \  jj\chi^{0}_{1}. 
\eea\
Figure \ref{fig:n2decay}(b)
shows the decay branching fractions of $\chi^{0}_{2}$ versus the wino
mass parameter $M_{2}$ for $M_{1}=120$ GeV.
For $M_{2} \lsim M_{1}$ where $\chi^{\pm}_{1}$  and $\chi^{0}_{1}$ are both wino-like and nearly degenerate, 
$\chi^{0}_{2}$ decays dominantly via charged currents.
For pure $\tilde{W}$ LSP, however, the mass difference between $\chi^{\pm}_{1}$ and $\chi^{0}_{1}$  
is only due to radiative correction and is of order $m_{\pi}$. The kinematically allowed decay is
$\chi^{\pm}_{1}\to \pi^{\pm}\chi^{0}_{1}$, and thus the NLSP can be long-lived. 
The thresholds reflect the kinematics due to the masses of $\tau$ and hadrons.
For $M_{2} > M_{1}$,  $\chi^{\pm}_{1}$  and $\chi^{0}_{2}$ are both wino-like and nearly degenerate. 
Then the $\chi^{0}_{2}$ decay to $\chi^{0}_{1}$ is strongly favored by kinematics.
%
%
\subsection{Signal Characteristics of Gluino Pair Production}

\begin{figure}[tb]
\includegraphics[scale=1.5,width=8.1cm]{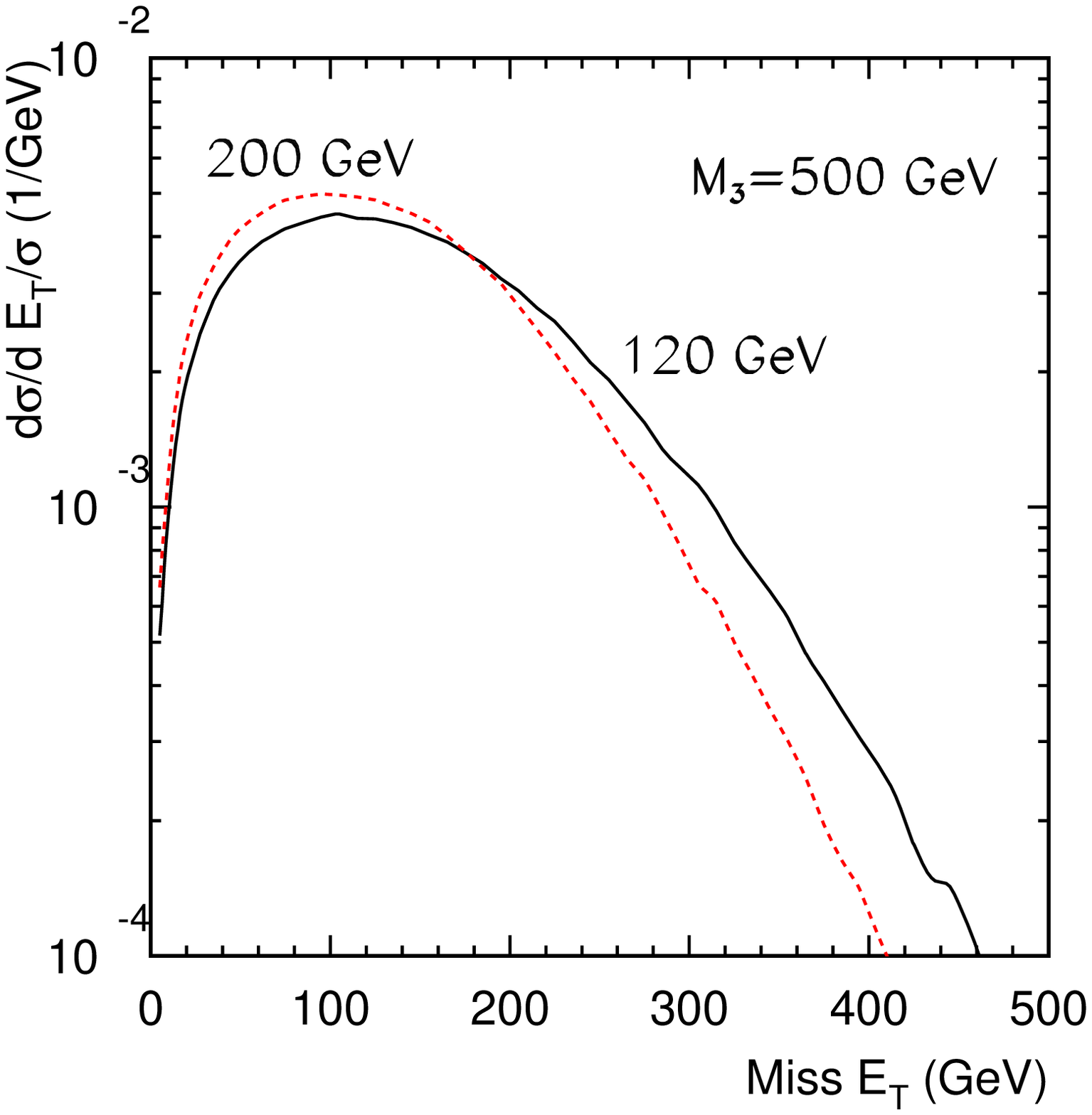}
\includegraphics[scale=1.5,width=8.1cm]{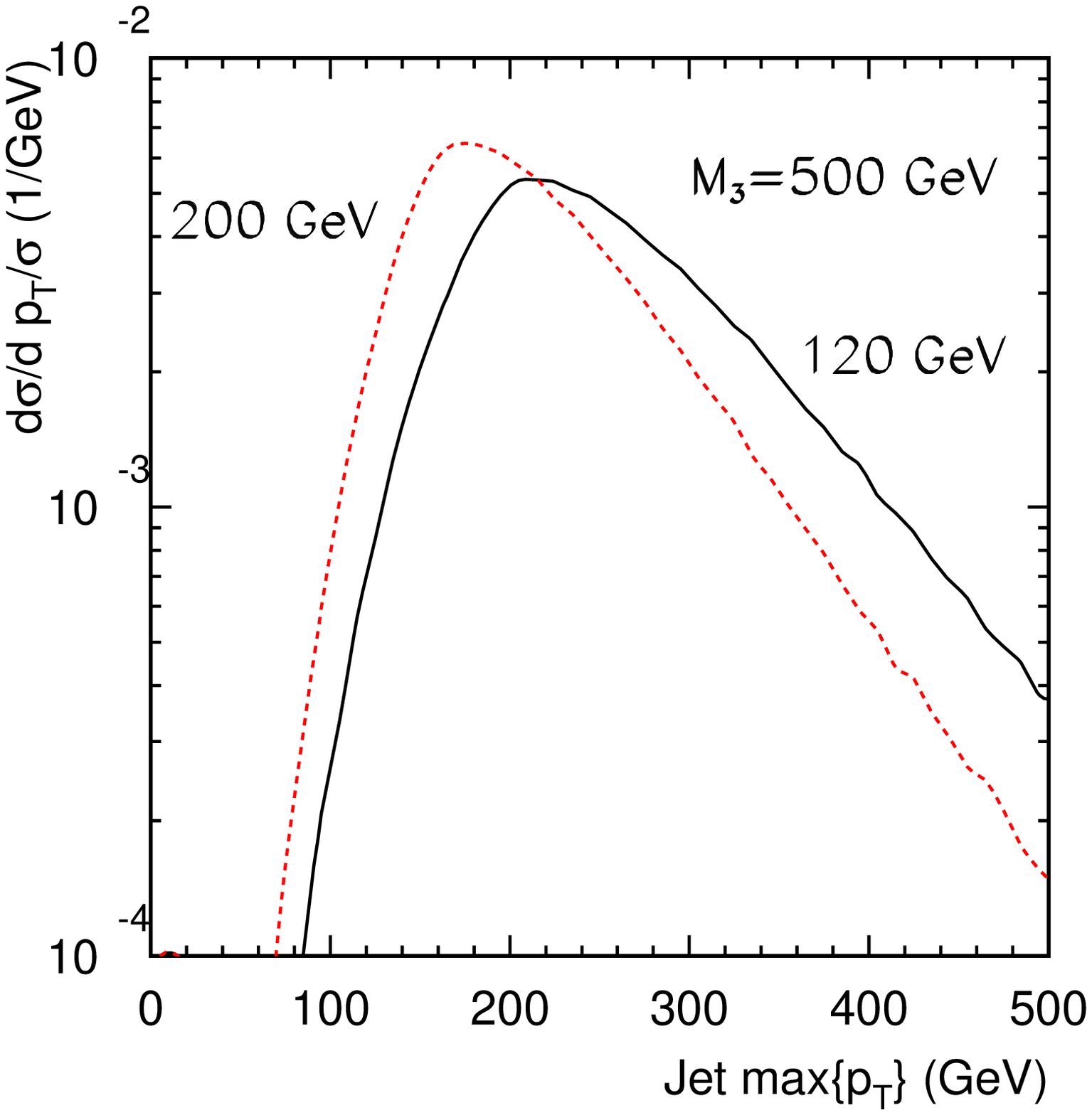}
\caption{Normalized transverse momentum distributions of the $\tilde g \tilde g$ signal 
at the 14 TeV LHC with $\mg =500$ GeV
for  (a) missing energy $\etmiss$ and (b) leading jet ${\rm max} \{p^j_T\}$, 
for  $M_1=M_2=120~\text{GeV}$ (solid curves), and $200~\text{GeV}$ 
(dashed curves). }
\label{fig:sigonly}
\end{figure}

As seen from the discussion in the previous section, 
gluino pairs usually lead to multiple jets with large missing energy, sometimes
accompanied by charged leptons 
($\ell^\pm =e^\pm, \mu^\pm$ for simplicity of the experimental observation). 
A pair of same-sign charged leptons, as a consequence of the Majorana nature of the gluino, 
is known to be a very important discovery channel at the LHC due to the low 
Standard Model background.  However, since we are mainly considering nearly degenerate
gauginos, the quarks and leptons from $\chi^{\pm}_{1}$  and $\chi^{0}_{2}$ decays will be
rather soft, and thus difficult to identify.  We now investigate and classify these signatures in detail. 

We first examine the jets plus missing transverse energy channel. We will  use several representative values of the mass parameters to illustrate  the basic kinematic features and design the basic event selection cuts. 
The  distributions of 
$\cancel{E}_T$ and the transverse momentum of a jet ($p_T^j$)
are determined mainly by the  difference between the gluino mass and the LSP~(NLSP) masses. Since we are considering the nearly degenerate $\tilde B - \tilde W $ scenario, we choose to study several values of $M_1$, and  vary $M_2 $ 
only by $30 - 50$ GeV around $M_1$, as in Sec.~II.A. 
Gluino mass $\miii$ both controls the production rate and affects the size of $\etmiss$ and $p_T^j$.  
 We begin by considering  a light gluino  $M_3 = 500$ GeV. 
Figure \ref{fig:sigonly} shows the distributions for the 
missing transverse energy and the hardest jet transverse momentum at the LHC 
for the two sets of parameter choices of Eqs.~(\ref{mixi}) and (\ref{mixii}).

As for our basic event selection, we first demand the signal to have a minimal 
missing transverse energy
\be
\cancel{E}_T> 100~\text{GeV}. 
\label{ETmiss}
\ee
The signal cross section after the $\etmiss$ requirement  is given by the dashed curve
in Fig.~\ref{fig:ggtot}.  We see that this selection becomes increasingly more
efficient for higher gluino masses.
Jets from heavy particle decays, such as  from gluinos, are typically harder than
the QCD jets in the SM. We thus require additional four jets in the events with
\be
p^j_T > 50\ {\rm GeV},\quad  |\eta_j|< 3.0,\quad  \Delta R_{jj} > 0.4, 
\quad \text{max}\{p^j_T\}>150~\text{GeV}.
\label{basic}
\ee

The high threshold in jet selection implies that 
the  hadronic decay of $\chi^{\pm}_{1}$ or $\chi^{0}_{2}$ as the leading channels 
will be largely invisible since the jets will be soft and will not pass the jet acceptance.

\begin{figure}[tb]
\includegraphics[scale=1,width=8cm]{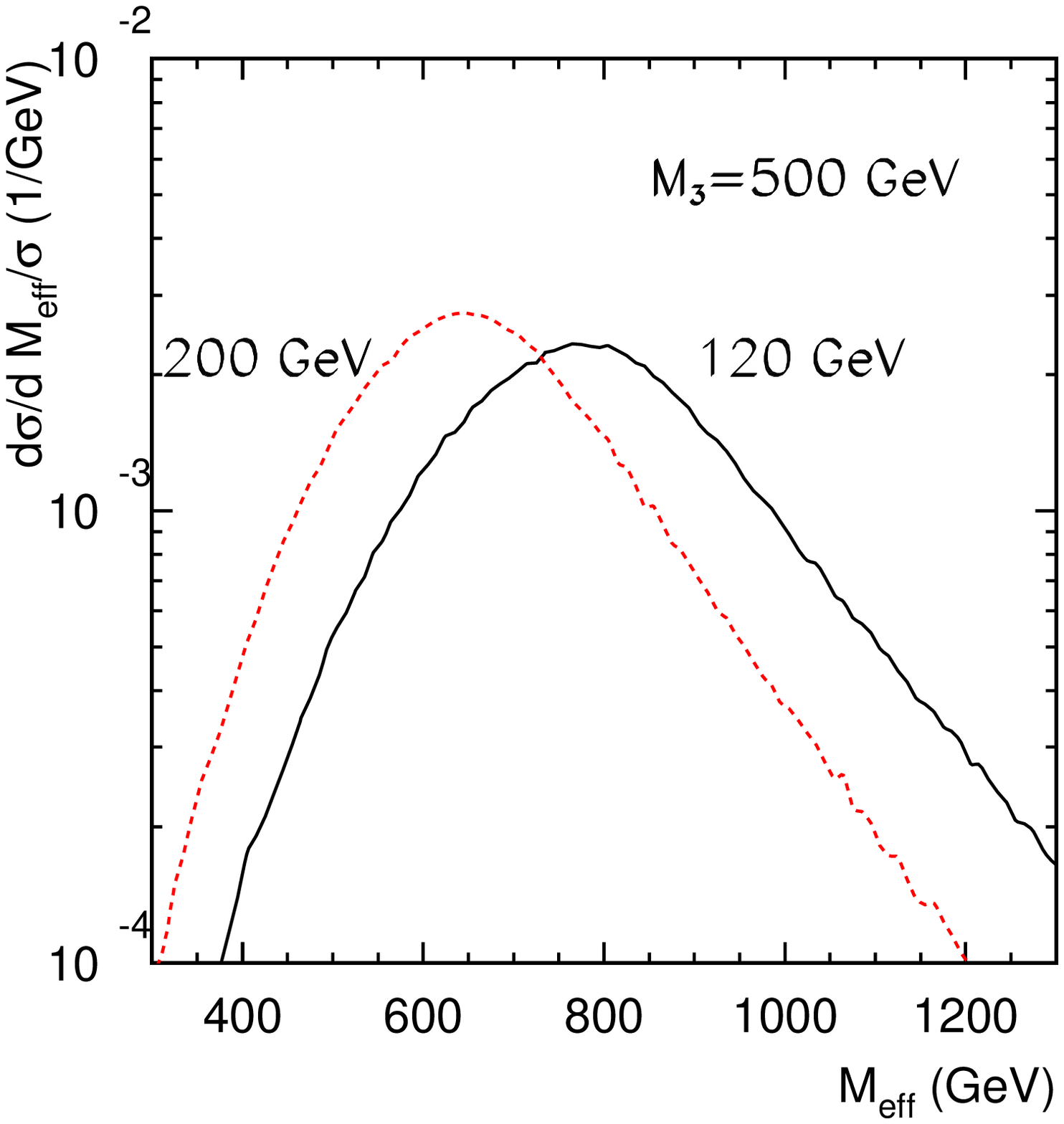}
\includegraphics[scale=1,width=8cm]{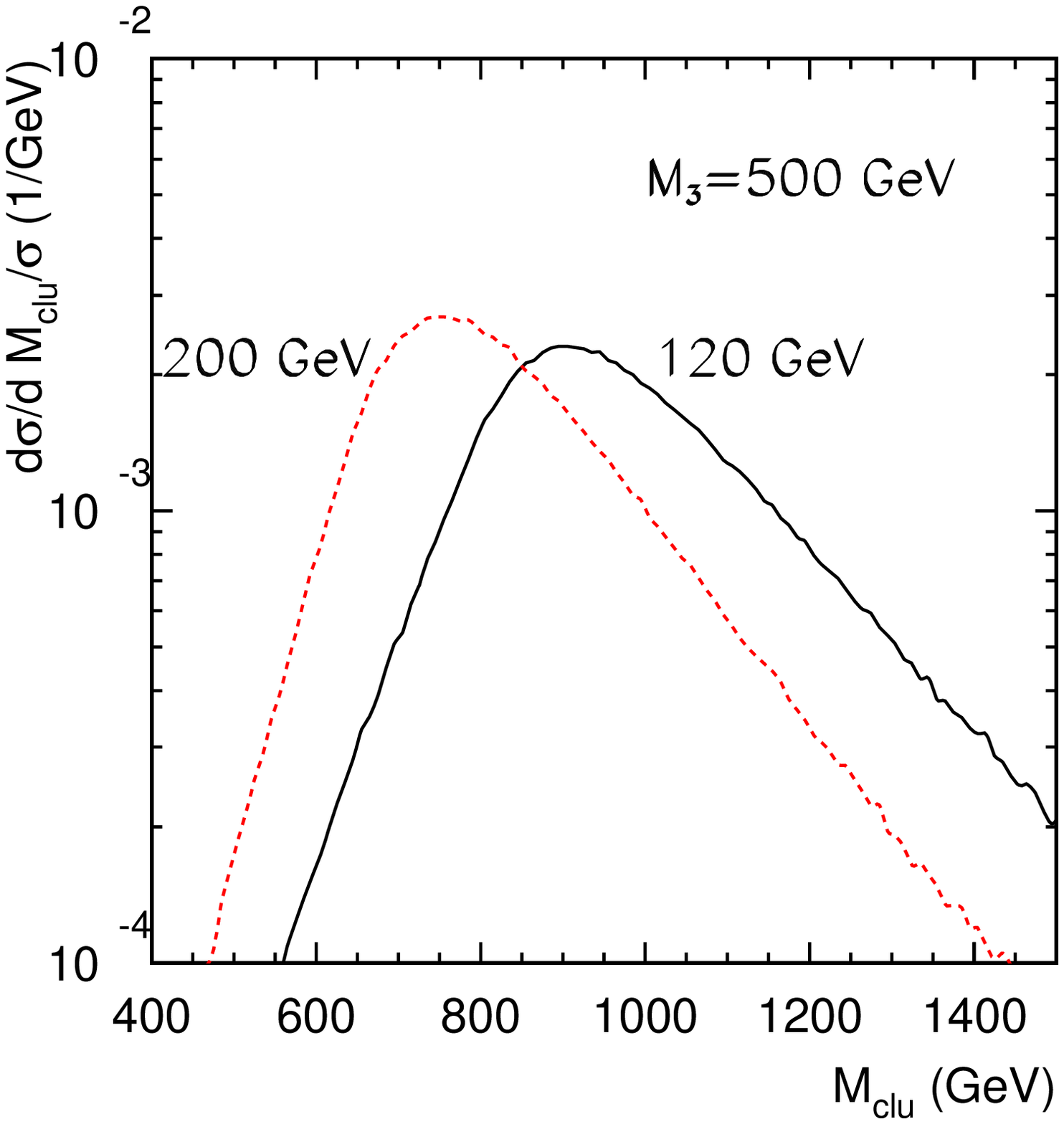}
\caption{Normalized  mass distributions of the $\tilde g \tilde g$ signal 
with $M_3=500$ GeV 
for (a) the effective transverse mass $M_{\rm eff}$ and (b) cluster transverse mass $M_{\rm cluster}$ 
for $M_1=M_2=120$ GeV (solid curves) and $M_1=M_2=200$ GeV (dashed curves). }
\label{fig:meff}
\end{figure}

In addition to the $\etmiss$ discussed above, some global mass variables provide a good 
measure for the energy scale in the case of heavy particle production. Typical examples of such variables include  the  ``effective transverse mass" and the ``cluster  transverse mass" defined as 
\bea
\nonumber
M_{\rm eff} = \sum_j |p_T^j | + \etmiss ,\quad \ 
M_{\rm cluster} =  \sqrt{M^2_{C}+( \sum_j \vec p^j_T)^2}+\cancel{E}_T ,
\nonumber
\eea
where the sum runs over all observable objects (jets, leptons etc.), and
$M_C$ is the invariant mass of the system of observed objects in the final state. 
Note that the effective mass is just the transverse mass defined by the massless objects
(jets, leptons etc.) and missing energy in a whole event. The cluster transverse mass 
is based on the grouped cluster of the observed objects.
We plot the distributions of the effective mass in Fig.~\ref{fig:meff}(a)   
and  of the transverse mass in Fig.~\ref{fig:meff}(b). The qualitative difference
with respect to the SM background is that
these two variables have broad peaks which is correlated with the mass difference  $\sim 2( \mg -\mlsp) $. 
We find it effective to impose an additional cut to further separate the signal
from backgrounds and suggest to adopt 
\begin{equation}
M_{\rm cluster} > 2 (\mg -\mlsp).
\label{eq:m}
\end{equation}
This cut is only meant to be qualitative. We do not assume to know the mass parameters,
but  some kinematical cuts 
should be optimized in realistic simulations for different masses of the gluino  and the LSP.

\begin{figure}[tb]
\includegraphics[scale=1.5,width=8.1cm]{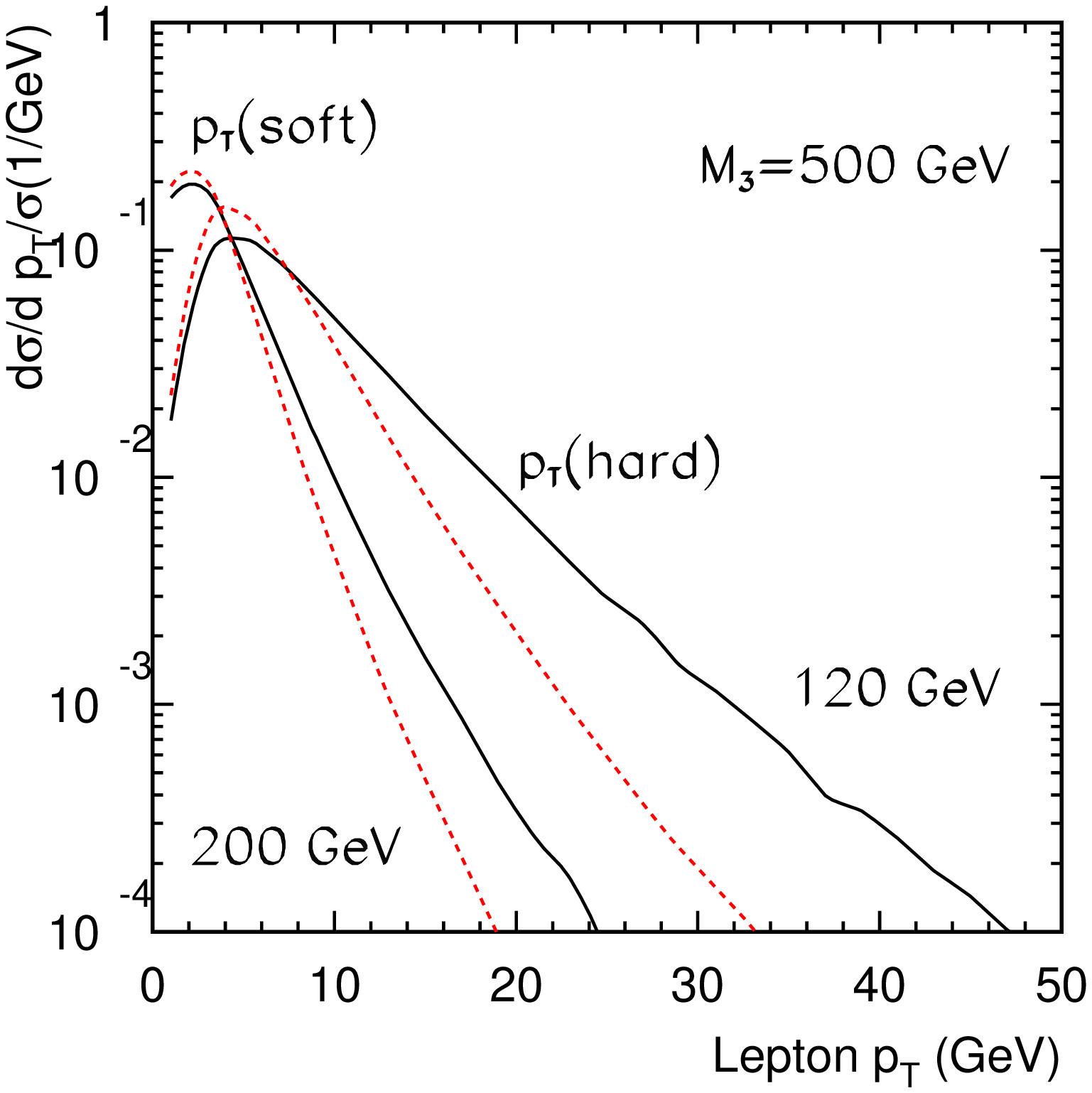}
\includegraphics[scale=1.5,width=8.1cm]{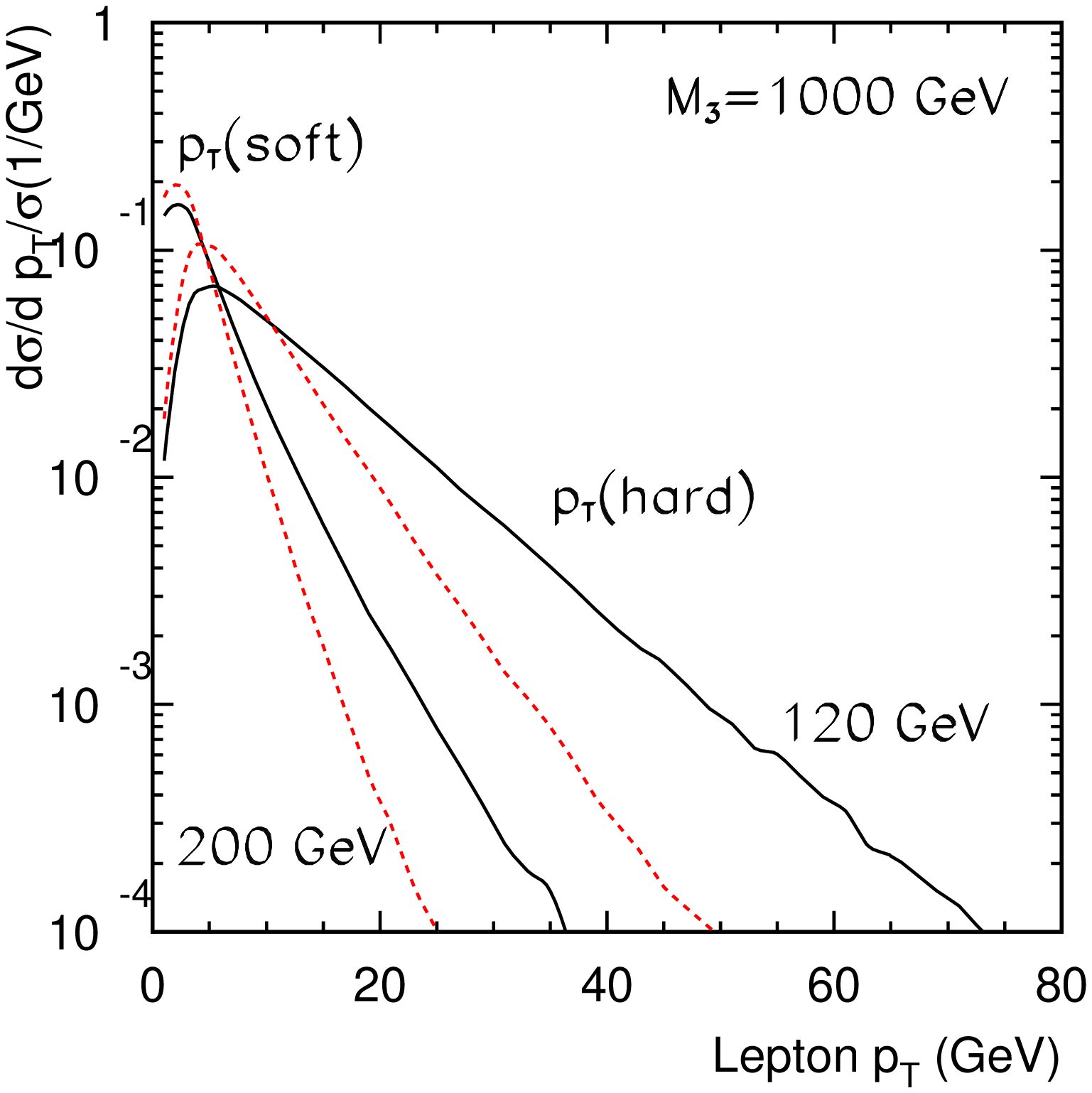}
\caption{Normalized transverse momentum distributions of the $\tilde g \tilde g$ signal 
for the soft leptons  with $\dm=8$ GeV, for (a) $\mg =500$ GeV and (b) $\mg =1000$ GeV,
for  $M_1=120$ (solid curves) and 200 GeV (dashed curves). }
\label{fig:gl}
\end{figure}

The gluino decay chains listed in Eq.~(\ref{eq:gg})  can often have charged leptons
in the final state. To understand the kinematical features of those leptons, 
we show the normalized transverse momentum distributions  in Fig.~\ref{fig:gl} 
for the softer and the harder leptons in events 
$\tilde{g}\tilde{g}\to 4j+\chi^+_1\chi^-_1\to 4j +2\ell+\cancel{E}_T$ with $\dm=8$ GeV,
for $\mi = 120$ GeV (solid curves) and 200 GeV (dashed curves). We see that a heavier 
LSP renders the $p_T^\ell$ spectrum softer. 
The harder spectrum of the leptons in  Fig.~\ref{fig:gl}(b) is obviously due to the boost 
effect from a heavier gluino. 

Including these leptons with moderate transverse momentum, $p^{\ell}_{T} \sim 10$ GeV as part of the signal identification 
can change significantly the search strategy. Instead of searching for those soft leptons at the trigger level, we envision looking for them  with off-line analyses. 
We begin  with a discussion of the importance of various channels in different regions of the parameter space.   
In our analysis, we use the following selection requirement for observing an isolated charged lepton 
(both electron and muon),  
\be
p^\ell_T > 10\ {\rm GeV},\quad |\eta_\ell|< 2.8,\quad
\Delta R_{j\ell},\ \Delta R_{\ell\ell}> 0.4.
\label{eq:l}
\ee

\begin{figure}[tb]
\includegraphics[scale=1.5,width=8cm]{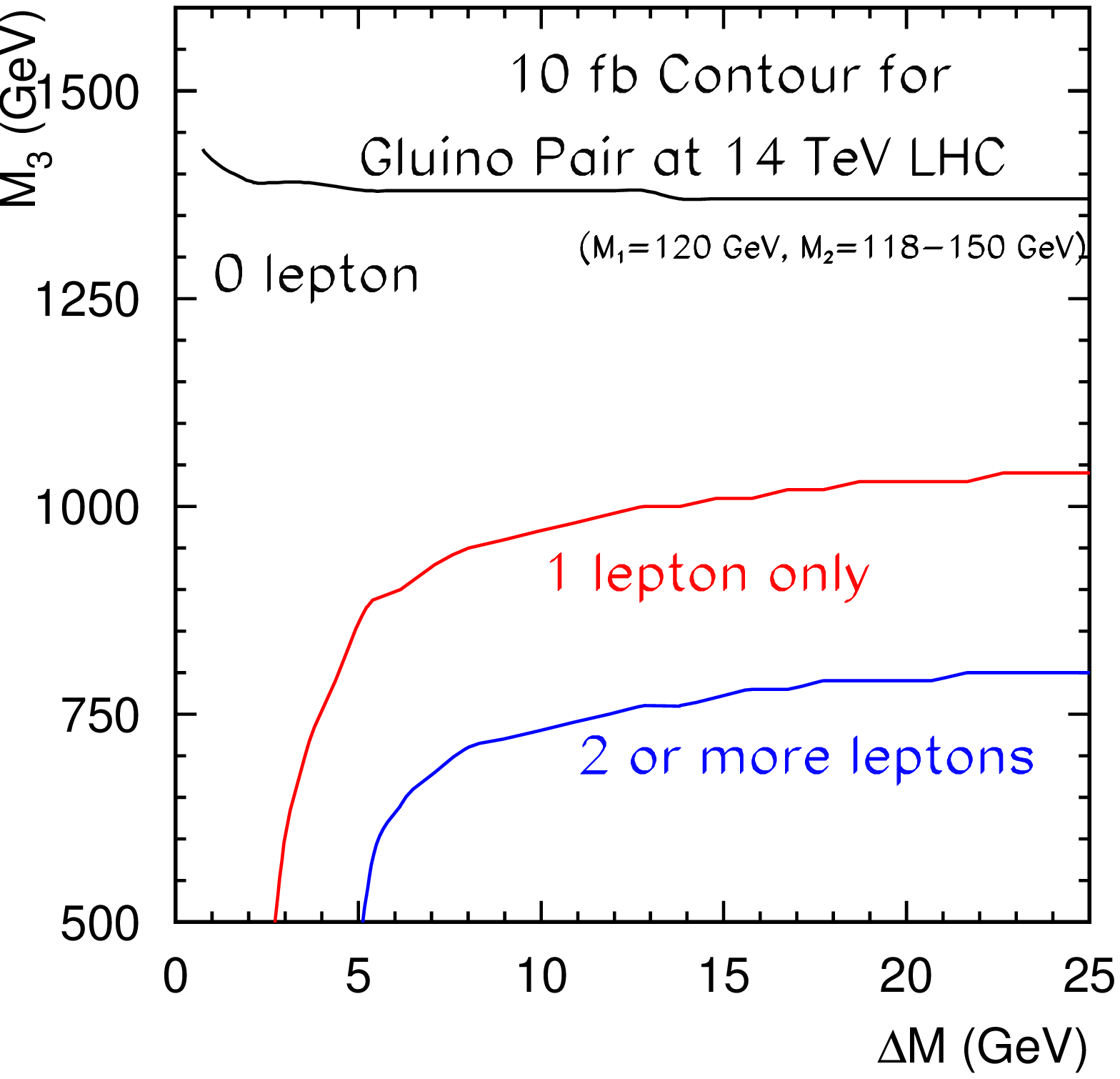}
\includegraphics[scale=1.5,width=8cm]{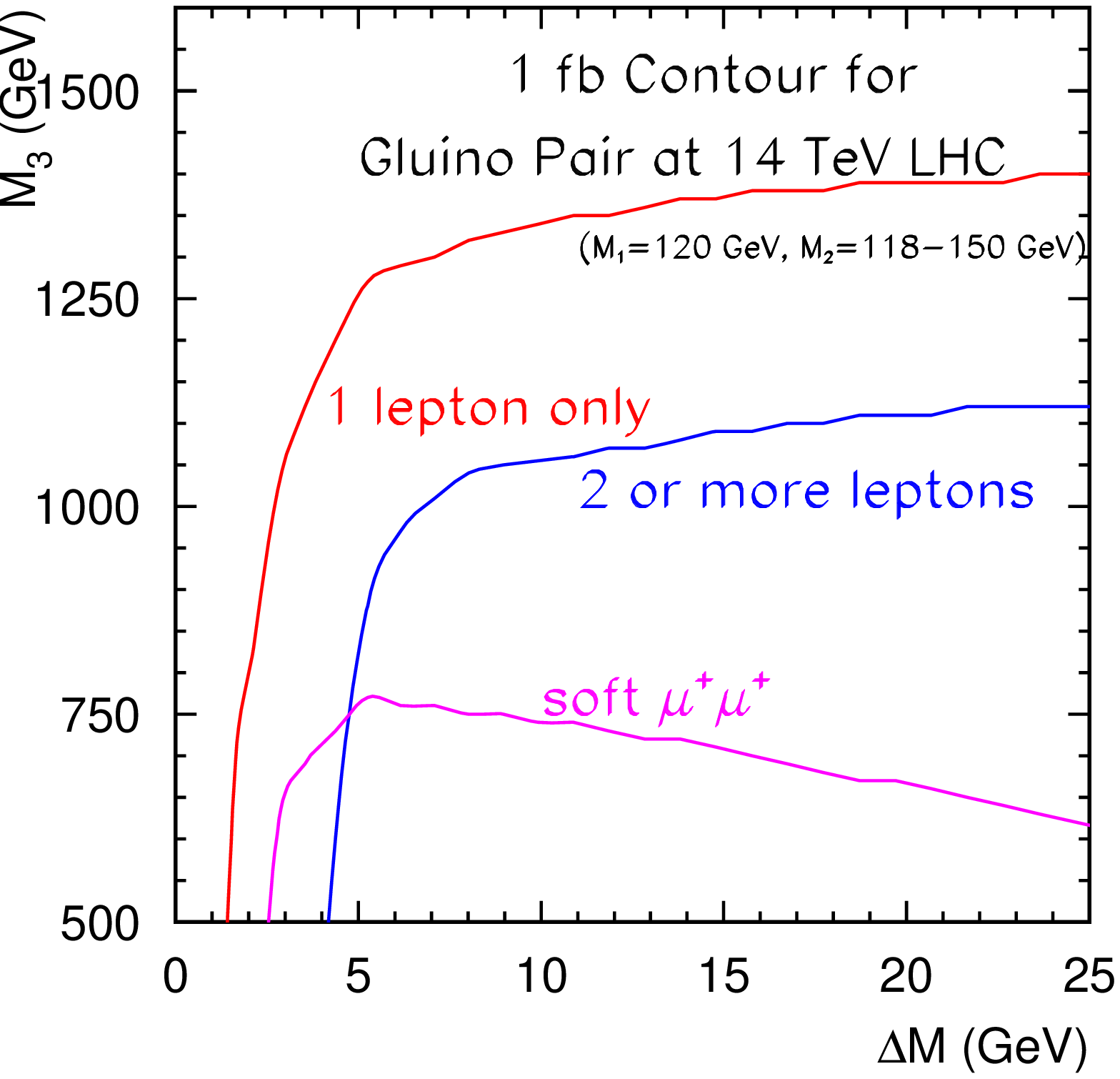}
\caption{Boundaries of regions with sizable cross section of the $\tilde g \tilde g$ signal 
for zero-lepton, 
one-lepton and two-lepton events from $\tilde{g}\tilde{g}\rightarrow 4j \chi_i\chi_j$,
in the plane of gluino mass versus the gaugino mass difference, with $M_1= 120$ GeV.  
The cuts of Eq.~(\ref{basic}) and Eq.~(\ref{eq:m})  have been imposed. 
In the region below each line, the rate is greater than 10 fb (a) and 1 fb (b) for the given channel.  
A 1 fb boundary for same sign soft dimuons, defined in Eq.~(\ref{eq:ptl}), is also included 
in (b) for later reference.  
}
\label{evt}
\end{figure}

Under the selection cuts in Eqs.~(\ref{ETmiss}), (\ref{basic}) and (\ref{eq:l}),
we plot the cross section contours of 10 fb and 1 fb in $\mg-\dm$ plane
as in Fig.~\ref{evt}, where we have used the gaugino parameters in Eq.~(\ref{mixi}).
In the region above one specific line, the rate for the corresponding 
final state is less than 1 fb and below the line the rate is larger. 
For example, zero lepton refers to  the final state where no lepton pass our selection
cuts in Eq.~(\ref{eq:l}). The zero lepton line, the boundary above
which the rate for zero lepton events drops below 1 fb, is decreasing with $\dm$
since we expect more event will have visible leptons for larger mass differences.  The one-lepton line is 1-lepton-only curve where 
there is only 1 lepton that passes our lepton selection cuts.
The two-lepton-or-more line bounds the region where at least two leptons
pass the lepton selection cuts.
The contours show the correlation between $\dm$ and $\mg$.
For the same gluino mass, a bigger mass difference $\dm$ leads to fewer zero-lepton
events. 
%
\begin{figure}[tb]
\includegraphics[scale=1,width=8.1cm]{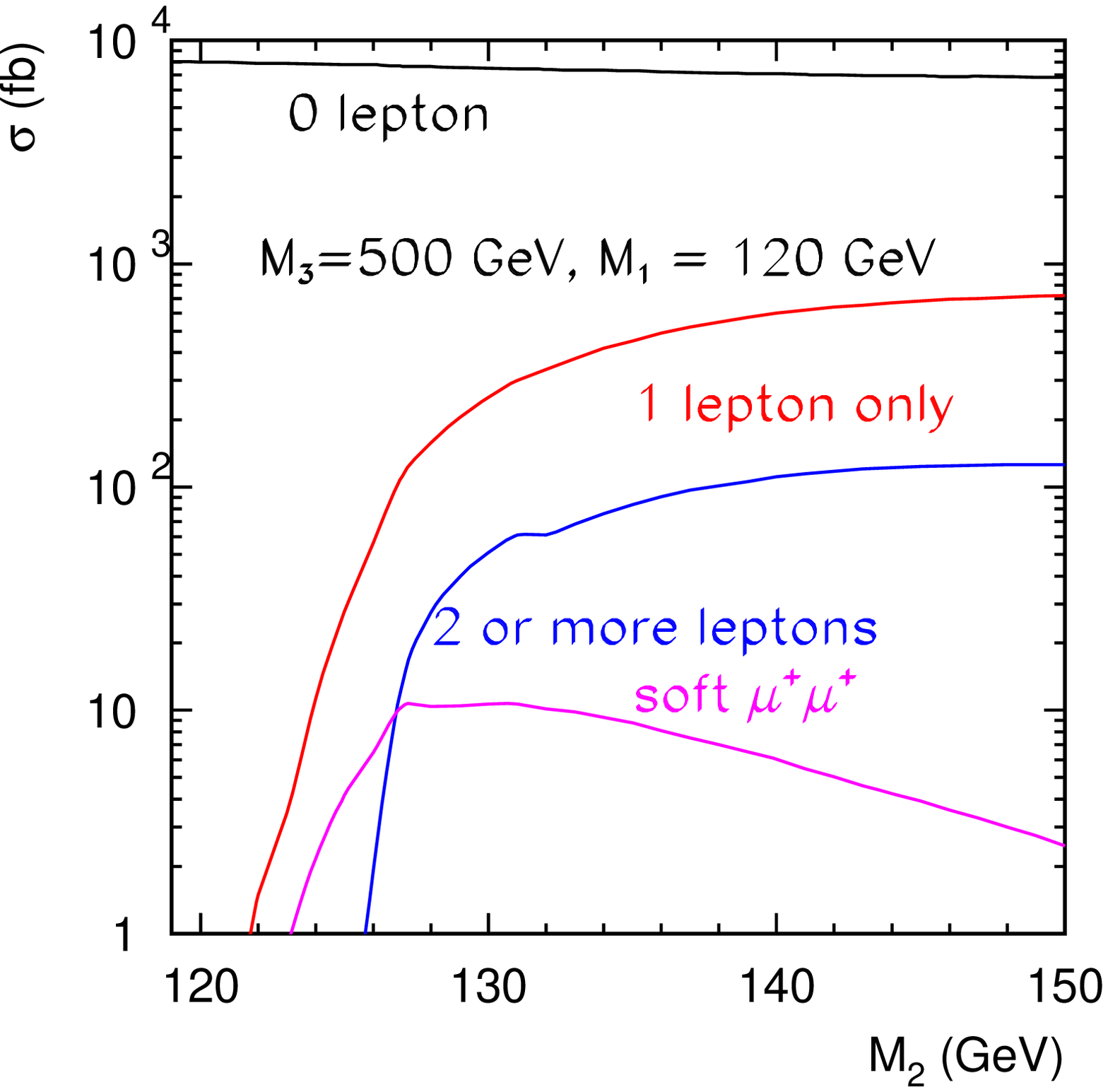}
\includegraphics[scale=1,width=8.1cm]{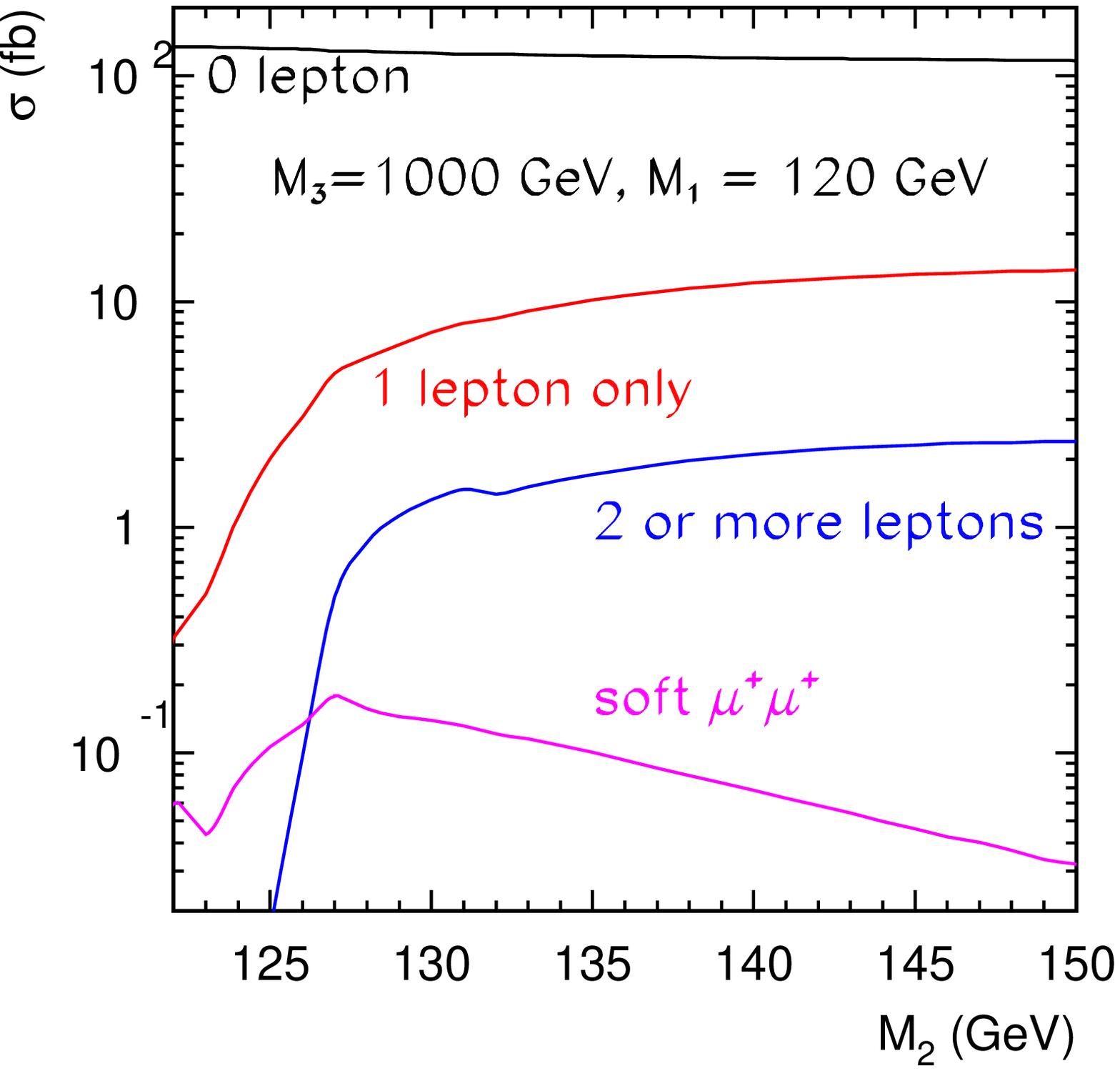}
\caption{Signal rates of the $\tilde g \tilde g$ final state
for zero lepton (solid), one lepton (dashed), and at least two leptons (dotted) for 
(a)  $M_3=500$ GeV and (b) 1000 GeV with $M_1=120$ GeV. The cuts of Eqs.~(\ref{basic}) and (\ref{eq:m})  
have been imposed. Cross sections 
for same sign soft dimuons, defined in Eq.~(\ref{eq:ptl}), are also included for later reference.}
\label{rate}
\end{figure}

The signal rates for two representative gluino masses  after imposing the cuts of Eqs.~(\ref{basic}) and (\ref{eq:m})   
are plotted in Fig.~\ref{rate} for different final states, 
$4j+\cancel{E}_T$ plus zero-lepton (solid curves), only one lepton (dashed), and 
at least two leptons (dotted) passing the cuts of Eq.~(\ref{eq:l}). As expected, 
the no-lepton case is an important channel for degenerate gauginos and
the rates for one-lepton and two-or-more-leptons are much smaller. 

\subsection{Observability of Jets$+\etmiss$ Signal}

The presence of the additional leptons  can potentially provide more handles in signal selection, as is well known when 
the mass splitting is sufficiently large \cite{Baer:1995va,atlasTDR}. 
However, we would like to emphasize that these leptons under consideration 
are not that hard due to the nearly degenerate gauginos. 
Moreover, unlike some more favorable cases with on-shell sleptons as part of the cascade, the leptons are dominantly from off-shell 
$W/Z$ decays in the our case. Therefore, the signal rate is further suppressed by the leptonic branching fractions. 
Leptons from Standard Model $W/Z$ decays, although typically harder, $p_T^{\ell} \sim 20-40$ GeV, still pose serious background to these leptonic channels.  
Therefore, we first focus on channels which do not rely on identifying isolated hard leptons. 
The most obvious channel in this category is jets$+ \etmiss$.

\begin{table}[tb]
\begin{tabular}{ |c|| c| c| c || c|}
    \hline
 SM backgrnds &  Basic cuts &  $M_{\rm cluster}$ cut & $M_{\rm cluster}$ cut & 1-soft muon \\
(pb)  & Eqs.~(\ref{ETmiss}), (\ref{basic}) & $>750$~GeV & $>1750$~GeV & $M_{\rm cluster}$ \\ 
 & & & & $>750$~GeV, Eq.~(\ref{eq:ptl}) \\
     \hline
     \rule[5mm]{0mm}{0pt}
     $Z$+4-jets &  110 & 96 & 25.1 & $-$\\
     \hline
      $W$+4-jets with $W\to \ell \nu_{\ell}$ &  4.6 & 3.3 & 0.4 & 1.5 \\
      $W$+4-jets with $W\to \tau \nu_{\tau}\to\ell\nu_{\ell}\nu_{\tau}\overline{\nu_{\tau}}$ &  5.1 & 3.6 & 0.4 & 1.1 \\
      $W$+4-jets with $W\to \tau\nu_{\tau}\to \nu_{\tau}\overline{\nu_{\tau}}+$ pions &  9.3 & 6.8 & 1.0 & $-$ \\
       \hline
      $t\bar{t}$ with $W\to \ell \nu_{\ell}$ (fb) &  83 & 33 & 0.6 & 14 \\
      $t\bar{t}$ with $W\to \tau \nu\to\ell\nu_{\ell}\nu_{\tau}\overline{\nu_{\tau}}$ (fb) &  107 & 38 & 0.7 & 11 \\
      $t\bar{t}$ with $W\to \tau \nu\to \nu_{\tau}\overline{\nu_{\tau}}+$ pions (fb) &  380 & 120 & 4 & $-$ \\     \hline
 Sum ~~~ (pb) &  129.5  & 109.9 & 26.9 & 2.6 \\
 \hline
\hline
  \end{tabular}
\caption{Cross sections after consecutive cuts for the leading SM background $4j+\cancel{E}_T$ as well as $4j+\cancel{E}_T+1$
soft muon as defined in Eq.~(\ref{eq:ptl}). 
We impose the cuts of Eqs.~(\ref{basic}) and (\ref{eq:m}),  and veto the events with leptons 
satisfying Eq.~(\ref{eq:l}).  The rate of soft muon is obtained after requiring $M_{\rm cluster}>750$~GeV.  }
\label{tab:gg}
\end{table}

The leading SM backgrounds for this signal are from electroweak gauge bosons plus 
QCD jets, as well as $t \bar t$ production
\bea
pp &\to& Z+4\ {\rm jets\ with}\ Z\to \nu\bar\nu\nonumber\\
     & \to &  W+4\ {\rm jets} ~({\rm including}~~ t\bar{t}\to W+4\ {\rm jets})~~{\rm with}~~W\to \ell \nu,\ \tau\nu_{\tau}, 
\eea
where the charged lepton $\ell$ from the $W$ and $\tau$ leptonic decays are below the lepton acceptance 
in Eq.~(\ref{eq:l}), hence missing from detection. 
Given our hard jet selection 
cut is $p^{j}_{T}> 50$~GeV, the hadronic $\tau$'s are very unlikely to be counted as a jet, 
leading to the dominant contribution of the visible $W$-decay channels. 
Since the $\ell$ from leptonic $\tau$ decay ($\tau\to\ell\nu_{\tau}\nu_{\ell}$)
are typically much softer than the $\ell$ from $W$ decay, even though the leptonic decay BR of $\tau$ is only
about  $35\%$, the leptonic $\tau$ contribution to invisible mode is as large as the contribution due to $W\to\ell\nu$ channel.
The basic cuts in  Eqs.~(\ref{ETmiss}) and (\ref{basic})  
already substantially reduce the SM backgrounds. 
The leading SM backgrounds of $4j+\cancel{E}_T$ are summarize in Table \ref{tab:gg} with consecutive acceptance cuts. 

\begin{figure}[tb]
\includegraphics[scale=1,width=8.1cm]{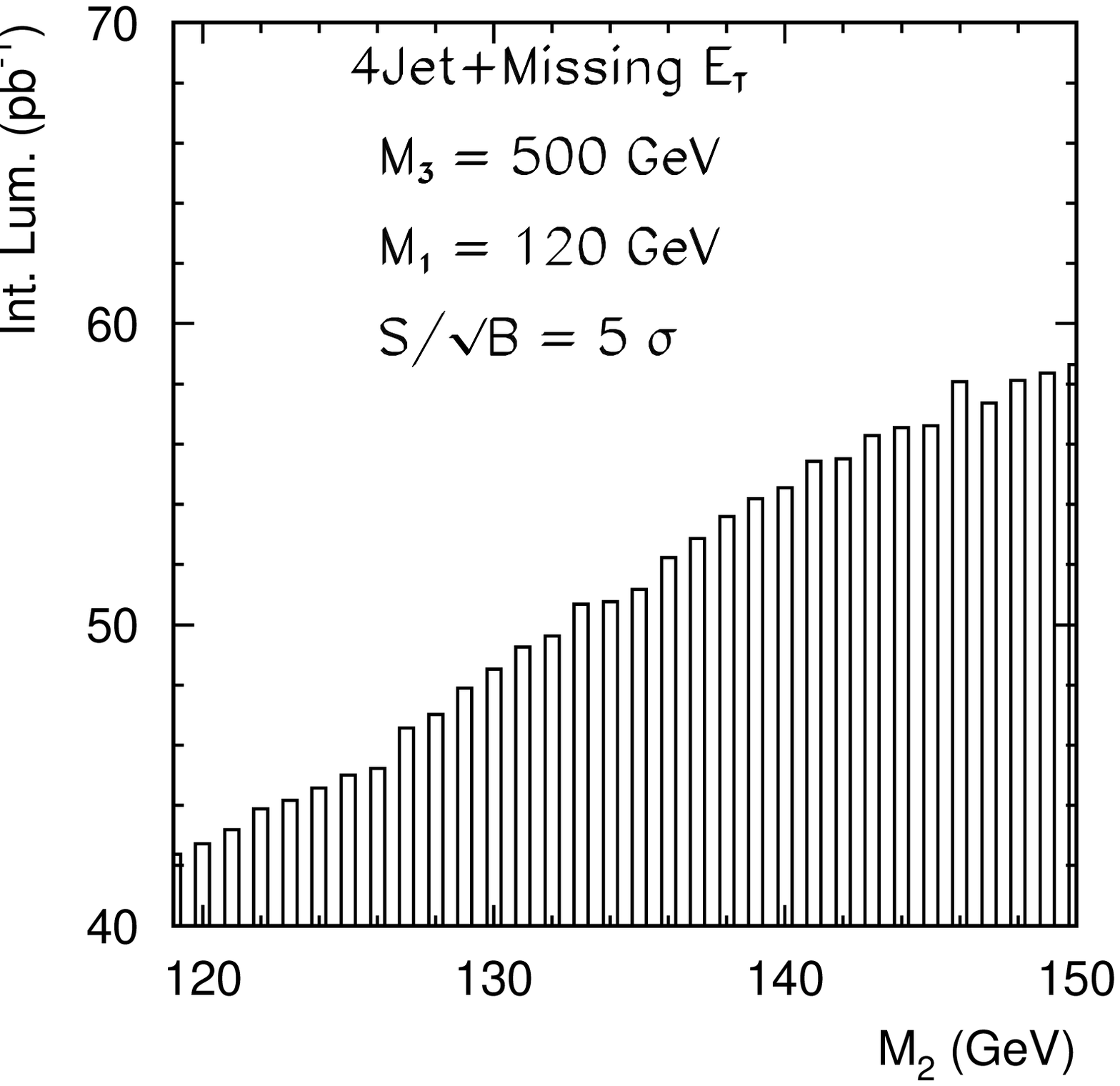}
\includegraphics[scale=1,width=8.1cm]{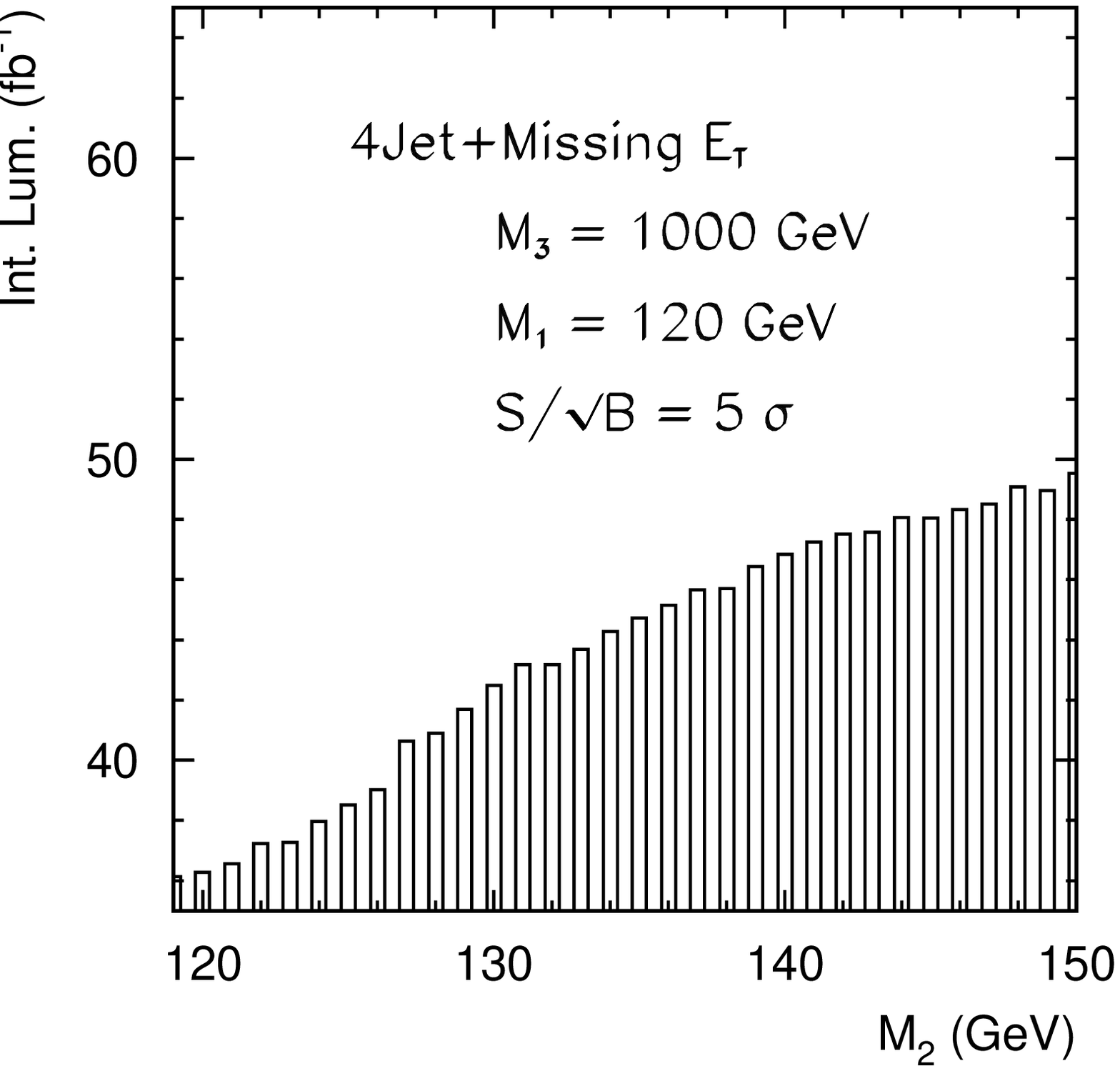}
\caption{Integrated luminosity needed for a statistical significance $S/\sqrt B=5$ of the $\tilde g \tilde g$ signal 
versus $M_2$ in the $\etmiss+$jets channel 
for $M_1=120$ GeV, and two representative gluino masses  $M_3=500$ GeV (left panel) and $M_3=1$ TeV (right panel) . 
The cuts of Eqs.~(\ref{basic}) and (\ref{eq:m}) have been imposed, and the events with a least 1 harder lepton 
satisfying Eq.~(\ref{eq:l}) have been vetoed.  }
\label{signif}
\end{figure}
We present our signal analyses for two representative scenarios with $M_3 = 500 $ GeV and $M_3=1$ TeV. 
The  mass splitting between the nearly degenerate gaugino states is varied. We have imposed the cuts of 
Eqs.~(\ref{basic}) and (\ref{eq:m}). In addition, to suppressed the large Standard Model backgrounds  with harder leptons 
from $W/Z$ decays, we veto events with leptons satisfying Eq.~(\ref{eq:l}). Combining with the background studies
above, the integrated luminosity needed to reach $5\sigma$ statistical significance ($S/\sqrt B=5$) 
for the $4j+\cancel{E}_T$ channel is shown versus in Fig.~\ref{signif}. 
The integrated luminosity needed to reach this sensitivity for the multi-jet plus missing energy signal  is 
 about 50  pb$^{-1}$ or 50  fb$^{-1}$ for a gluino mass of 500 GeV or 1000 GeV, respectively. 
We conclude that jets$+ \etmiss$ provides a promising channel for discovering supersymmetry in the case of nearly 
degenerate gauginos, regardless the presence of charged leptons or not. 
This should not be a surprise given the similar conclusions in the literature for the case of non-degenerate 
gauginos \cite{Baer:1995nq,atlasTDR,ETmissPapers}. 

\subsection{Soft Leptons in Jets$+\etmiss$ Events}

 \begin{figure}[tb]
\includegraphics[scale=1,width=8.12cm]{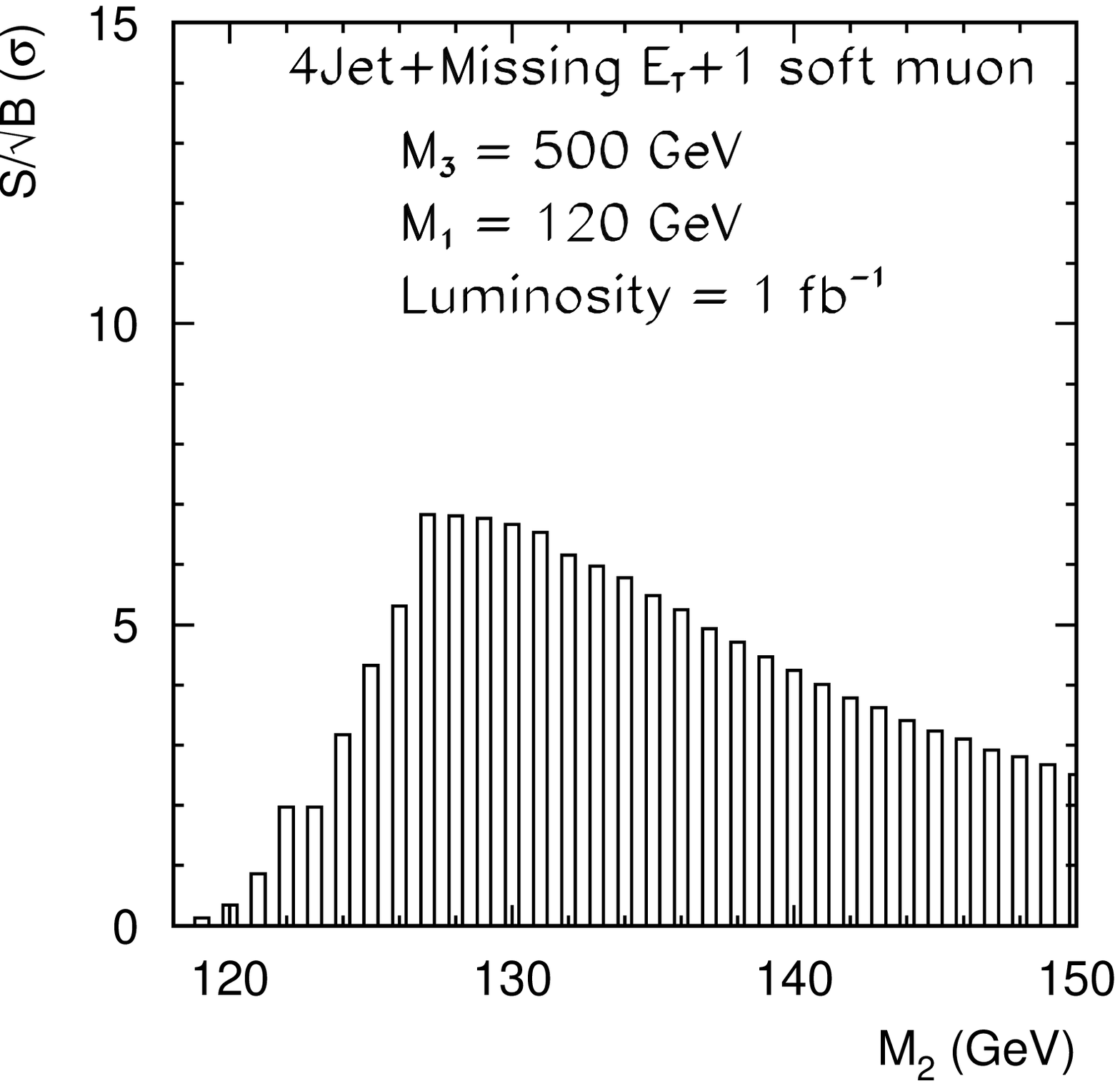}
\includegraphics[scale=1,width=8.12cm]{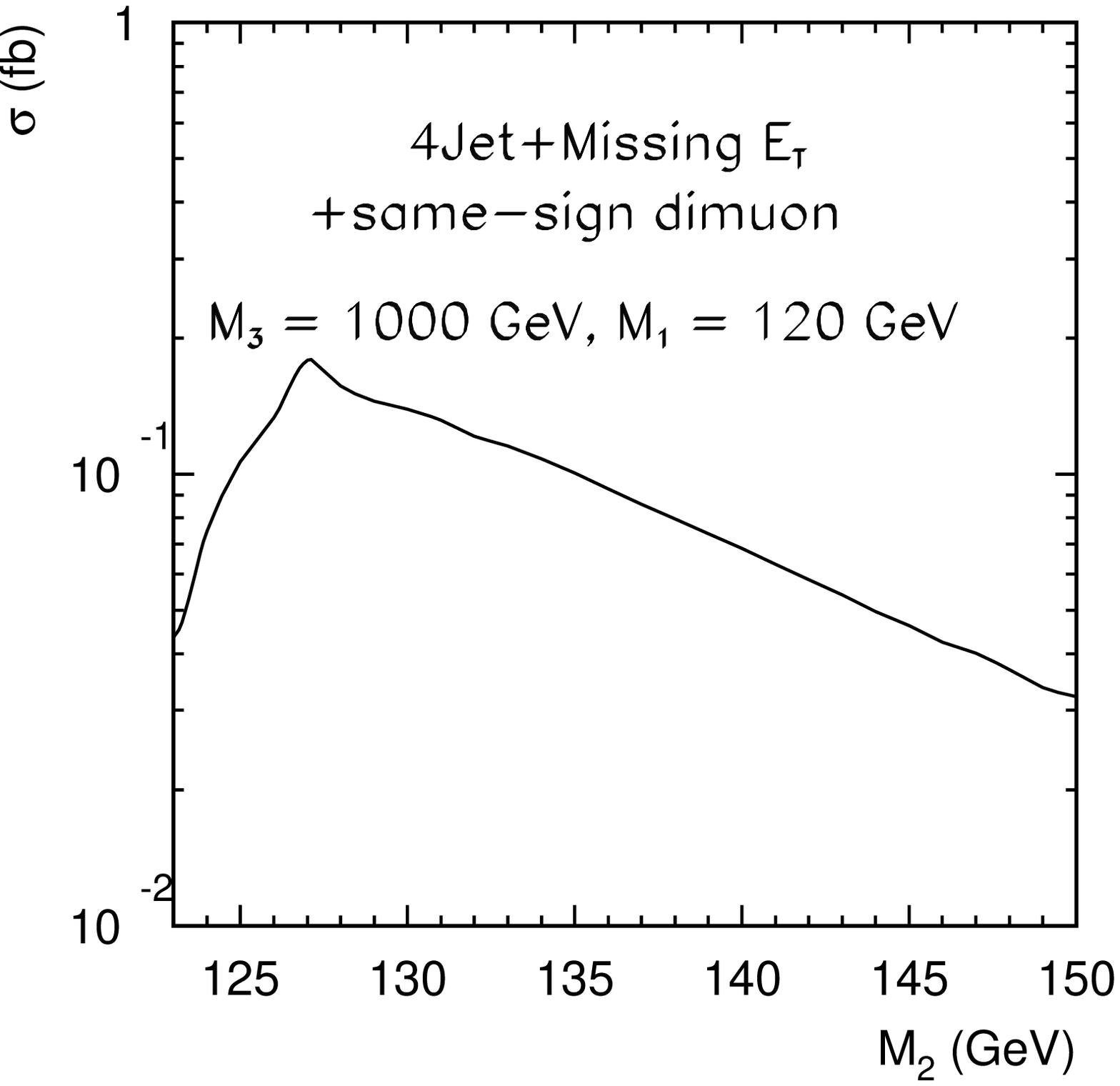}
\caption{ (a) Statistical significance $S/\sqrt{B}$ of the $\tilde g \tilde g$ signal 
for 1~fb$^{-1}$ luminosity for 
$4j+\cancel{E}_{T}+\mu^{\pm}$ events with $M_{3}=500~$GeV.
(b) Soft muon signal cross sections for $4j+\cancel{E}_{T}+\mu^{\pm}\mu^{\pm}$
with $\mg=$1 TeV.  }
\label{sigmuon}
\end{figure}

If some signal events of $4j+\cancel{E}_T$ type are discovered, it will become crucial to assess if
they indeed come from the SUSY prediction of nearly degenerate gauginos. 
Such an evidence could be inferred from the observation of isolated 
soft charged leptons produced in the decay chain 
$\chi_1^\pm ,\  \chi_2^0  \to \chi_1^0\  \ell^\pm$'s, namely from the events
\be
4~{\rm jets} +\etmiss + \ell^\pm_{\rm soft}.
\ee
To explore this possibility, we revisit the $p_T^\ell$ distributions in Fig.~\ref{fig:gl},
where the $p_T^\ell$ spectrum of the soft leptons is controlled by $\dm$. We see that in a large fraction of the leptonic events, the lepton is rather soft with $p_T^{\ell} \lsim 10 $ GeV.  
Therefore, we propose to look  for isolated soft muons in the kinematical region 
\begin{equation}
3\ {\gev} < p_T^\mu < 10\ {\gev},\quad |\eta_\mu | < 2.8,\quad \Delta R_\mu > 0.4.
\label{eq:ptl}
\end{equation}
The upper limit on $p_T^{\mu}$ is enforced by the lepton veto  described earlier in this section in order to suppress the background from leptonic decays of $W$ and $Z$.  
The background for the $4j+\cancel{E}_T+1$ isolated soft muon is shown in the last 
column of Table \ref{tab:gg}. 
The dominant backgrounds are $W+4$ jets and $t \bar{t}$ with $W\to\mu\nu_{\mu}$ and $W\to\tau\nu_{\tau}\to \mu\nu_{\tau}\nu_{\mu}\nu_{\tau}$. 
$Z+4$ jets gives negligible background in this case due to the absence of large $\etmiss$ in this channel. We compare this background with a typical signal with
$M_3=500$ GeV  in Fig.~\ref{sigmuon}(a) using the statistical significance $S/\sqrt{B}$ for 1~fb$^{-1}$ integrated luminosity data. We see this channel can be useful if the mass splitting is about $5 - 30$ GeV. 
The signal rate is decreasing for larger mass differences since fewer  events pass our hard lepton veto.

%
Given the encouraging results for an isolated soft lepton above, we are thus motivated to consider two like-sign soft muons  as specified in Eq.~(\ref{eq:ptl}) in the final state 
\be
4~{\rm jets} +\etmiss + \mu^\pm_{\rm soft}\  \mu^\pm_{\rm soft}.
\label{eq:two}
\ee
This class of events can help to establish the Majorana nature of the gluinos \cite{Barnett:1993ea}. 
The leading irreducible background turns out to come from 
\be
t\bar{t}W^\pm \to b\bar b,\ 2j,\ \mu^\pm \mu^\pm+\etmiss.
\ee
After the stringent acceptance cuts the background is suppressed to a negligible level,
as shown in Table \ref{ttbw}.  As expected, due to the requirement of an additional same sign lepton,  this rather clean 
signal suffers from low rate  as plotted in  Fig.~\ref{sigmuon}(b), 
and higher luminosity would be needed for observation  of the signal.  

\begin{table}[tb]
\begin{tabular}{ |c| | c| c| c|}
    \hline
 Background &  Basic cuts & $M_{\rm cluster}$ cut &  2 same-sign soft muons \\
(fb)  & Eqs.~(\ref{ETmiss}), (\ref{basic}) & $>1750$~GeV & Eq.~(\ref{eq:ptl}) \\
     \hline
     \rule[5mm]{0mm}{0pt}
       $t\bar{t} W$ & $0.18$ &  $1.2\times 10^{-3}$ & $<10^{-4}$ \\
\hline
  \end{tabular}
\caption{Cross section rates with consecutive cuts for the leading SM background 
$t\bar{t} W$ to the signal  events of Eq.~(\ref{eq:two})}.
\label{ttbw}
\end{table}

In the study of soft lepton signals, 
we have only focused on the possibilities of observing the soft muons, with the expectation that it is easier to identify than a soft electron with similar $p_T$. Soft electrons can be included in the analysis by properly taking into account the experimental efficiency and fake rates. The resulting reach can be obtained by properly scaling our results. 

\subsection{Gluino Signal and SUSY Mass Parameters}

As shown in Figs.~\ref{fig:sigonly},  \ref{fig:meff} and \ref{fig:gl}, the global features of the kinematical 
distributions of the leptons and jets carry crucial information about the SUSY masses.  
The heavier the gluino is, the harder the kinematical distributions are,
while the heavier the LSP is, the softer the  distributions are. 
The experimental observables are governed by three mass parameters, $\mg$, $\mlsp$, and $\Delta M$. 
The gluino mass $\mg$ controls the signal production rate, while 
the mass difference $\mg - \mlsp$ determines the overall kinematical scale.
More precisely, the key features of the $M_{\rm eff}$ and $M_{\rm cluster}$ distributions, such as the peak and the average,  
are strongly correlated with the mass difference.  Other transverse variables display a similar correlation. 
The precise form of such a correlation can be obtained from careful simulation. Therefore, even with additional uncertainties from higher order corrections and experimental resolutions, a fit to these distributions can provide a useful measurement of $\mg - \mlsp$. 
If the gluino mass can be approximately obtained from other means, such as from
the total cross section measurements within a given theoretical model, 
then a first estimate of $M_{\rm LSP}$ can be extracted.

The most important parameter  to characterize the nature of nearly degenerate gauginos is $\Delta M$, 
and we have studied its effects in detail in this section. This parameter sets the kinematical scale for the NLSP decay
products and thus largely determines the interplay among the observed events with soft leptons/jets or not. 
Furthermore, a secondary parameter $\mg - \mlsp$ may be inferred as well. 
For fixed $\Delta M$,  
having a smaller $\mg -\mlsp$ will lead to softer jets and leptons, hence change the signal ratio of different class of events. 
Based on the jet selection cuts employed here, we expect the jets+$\not{\!\!E}_T$ channel can be effective  until $\mg - \mlsp \sim 100 $ GeV.
The effectiveness of the soft lepton channels with different $\mg-\mlsp$ can be estimated from our illustrative points, Fig.~\ref{evt}, and properly taking into account the boost effect.

Running at a lower energy $E_{CM}=7$ TeV obviously reduces the reach. 
The effect on the gluino channel is mainly from the reduction of the gluino production rate, 
shown in the right panel of Fig.~\ref{fig:ggtot}(b). Therefore, approximately,  we can rescale the reach accordingly.

While we have seen that gluino decays provide a promising channel to study the scenario with nearly degenerate gauginos, it is nonetheless important to consider other, more model-independent, channels. This leads us to explore the electroweak production of gaugino pairs.

\section{Gaugino Pair Production plus a jet: Mono-jet + $\etmiss$}

With or without any observable contribution from the gluino pair production 
of Eq.~(\ref{gluinos}), we should also consider the electroweak  gaugino-pair production in Eq.~(\ref{ewpp}). 
Whenever the final-state leptons are too soft, which is often the case in the nearly degenerate gaugino scenario,  we
are forced to consider the pair production processes in association with a hard QCD jet with large transverse 
energy to trigger on. 
This is the most model-independent WIMP (weakly-interacting massive particles) production channel,
common to many dark matter models. 
By kinematical crossing, this production mechanism is also related to the direct detection  processes
for the WIMP. 
%

 Discovery potential of a similar signal at the LHC has been studied in the focused point scenario \cite{Feng:1999zg,Baer:2005ky}, wino LSP scenario \cite{monojet}, and other variety of  scenarios \cite{monojet2}. 
 Recently, search of dark matter in the same final state at the Tevatron and the LHC has been studied in Ref.~\cite{Beltran:2010ww}.
The interaction of gauginos with the SM quarks under our current consideration,
both weakly coupled and without heavy intermediate state,  cannot be modeled in this formulation.

%

Notice that in the scenario with a pure wino LSP, such in AMSB \cite{amsb}, the
lightest chargino has a long lifetime. It will leave charge tracks
which give rise to unique signals  \cite{amsb_pheno,monojet}. A study of this class of signal  from a general class of new physics states have been carried out recently \cite{Buckley:2009kv}. We will not discuss further this
well studied scenario further in this article.

To effectively separate the signal from SM backgrounds, we  choose to impose an acceptance  
cut on the missing  transverse energy 
\be
\etmiss > 200\ {\gev}.
\label{eq:ETmcut}
\ee
Due to the mono-jet nature of the events, this  is equivalent to imposing a cut on the jet. 

\begin{figure}[tb]
\includegraphics[scale=1.0,width=9.5cm]{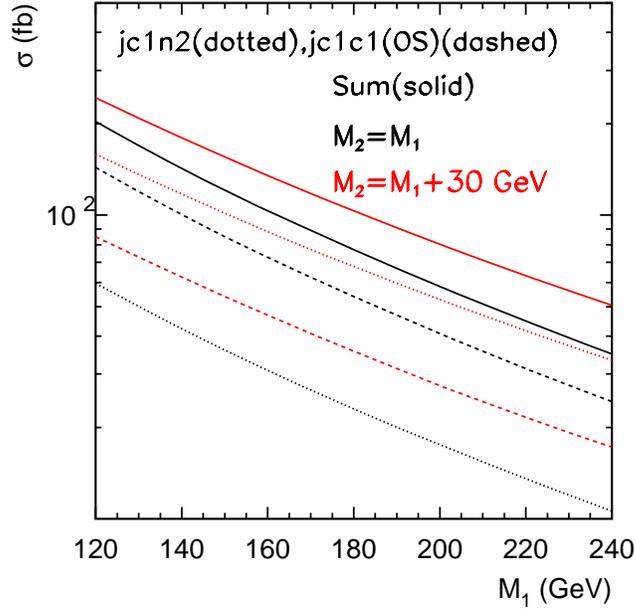}
\caption{
Total cross sections of the DY gaugino pair production 
versus $M_1$  for $M_2=M_1,\ M_1+30$ GeV and $\mu\to \infty$ 
where the dotted line refers to mono-jet+$\chi^\pm_1\chi^0_2$
(labelled as ``jc1n2''), the dashed line to  mono-jet+$\chi^+_1\chi^-_1$ (``jc+c-''), and the solid line to the sum.
The basic selection cut of Eq.~(\ref{eq:ETmcut}) has been imposed. }
\label{fig:m2mu}
\end{figure}

We first illustrate the variation of the signal rates with different choices of the SUSY parameters. 
We plot the total cross sections versus $M_1$ in Fig.~\ref{fig:m2mu}\footnote{Here and henceforth, we
also use ``ci (nj)'' to denote the i$^{th}$ chargino (the j$^{th}$ neutralino).}
for $M_2=M_1,\ M_1+30$ GeV, $\tan\beta=5$ and $\mu\to \infty$, 
with the basic selection cut of Eq.~(\ref{eq:ETmcut}). The cross section is typically
less than 0.2 pb. 
The production rate will be even more suppressed if $\chi^\pm_1$, $\chi^0_2$ are 
Higgsino-like as in mixed bino-Higgsino case. 

\subsection{Mono-jet Plus $\etmiss$ Signal}

\begin{figure}[tb]
\includegraphics[scale=0.3,width=9.5cm]{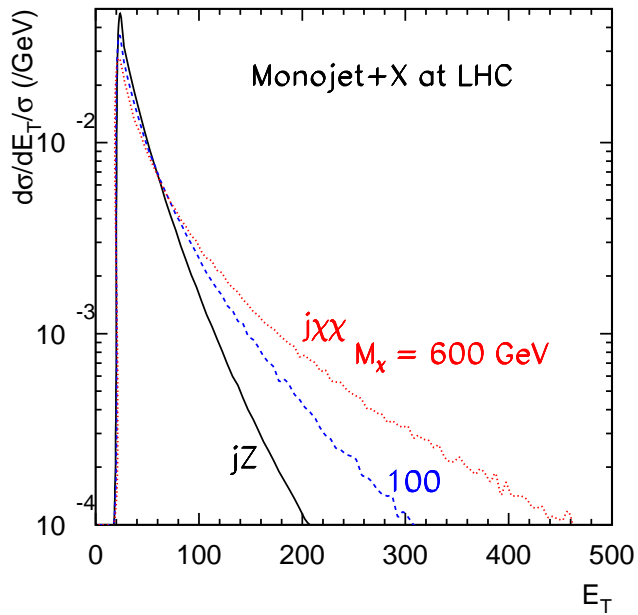}
\caption{Normalized transverse energy distributions of the DY gaugino pair production 
for the mono-jet$+ \cancel{E}_T$ channel 
from  $\chi^+_1 \chi^-_1 j$ with $M_{\chi^\pm_1}=100$ GeV (dashed), 
600 GeV (dotted) and the SM background $j Z$ (solid).}
\label{fig:ptjSB}
\end{figure}

The signal we are looking for is essentially an energetic mono-jet plus
large missing transverse  energy. In Fig.~\ref{fig:ptjSB},  we plot 
the normalized transverse energy distributions of the mono-jet for two extreme mass values 
of our interest $M_{\chi^\pm_1}=100$ GeV and 600 GeV.
One can see from the figure that heavier gauginos lead to a harder jet spectrum. 

The largest background in rate comes from QCD jets$+\etmiss$, where $\etmiss$ is 
due to the jet energy mis-measurement.
This background  falls very sharply at higher transverse energies and can be effectively suppressed
by the acceptance  cut of Eq.~(\ref{eq:ETmcut}). 
The leading irreducible background is from 
\be
Z+ 1~{\rm jet} \to \nu\bar\nu + 1~{\rm jet},
\ee
and there are also backgrounds 
\bea
W^\pm + 1~{\rm jet}~~ \text{with}~~ W^\pm\to \mu\nu_\mu ~~\text{or}~~ 
W^\pm \to \tau^\pm \nu_\tau\to\mu^\pm \nu_\mu\nu_\tau\nu_\tau 
\label{eq:W}
\eea
where the charged leptons escape from detection. Following the same argument for
$\tau$ hadronic decay, we also include the contribution from $\tau$ hadronic decay.
We tabulate these background rates with consecutive cuts in Table \ref{tab:ZW}. 
The total SM background sums to about 20 pb after the cuts, 
while the signal cross section for $M_1=120$ GeV is  about  0.2 pb. 
which may lead to a statistically significant signal.  For instance, with an integrated luminosity
of 15 fb$^{-1}$, this yields about a $5\sigma$ significance. 
However, due to the rather simple kinematics of the events, 
there is no distinctive feature in the shape of the distributions
between the signal and the leading background. 
Since the signal-to-background ratio (S/B)  is only at a $1\%$ level, 
the potentially large systematic uncertainties would render  the signal identification very
challenging if we only rely on the potential access in the total rate. 
Further refinement and improvement are possible such as exploiting the leptons in the events. 
We will next examine the events with soft muons. 

\begin{table}[tb]
\begin{tabular}{ |c| | c| c| c|}
    \hline
 Background &  Basic cut$+$lepton veto &  1 soft muon  \\
(pb)  & Eq.~(\ref{eq:ETmcut}) and $p^\ell_T< 10$~GeV & Eq.~(\ref{eq:ptl}) \\
     \hline
     \rule[5mm]{0mm}{0pt}
       $\nu\bar\nu + 1\ {\rm jet}$ & 13  & $-$ \\
       $\ell^{\pm} \nu + 1\ {\rm jet}$ &    2.2 &  0.42\\
       $\tau^{\pm} \nu + 1\ {\rm jet}$ with $\tau\to\ell\nu\nu$  & 1.5 & 0.38 \\
       $\tau^{\pm}\nu+1\ {\rm jet}$ with $\tau\to \nu+$ pions & 3.5 & $-$\\
\hline
  \end{tabular}
\caption{Cross section rates with consecutive cuts for the SM background 
to the  mono-jet+$\etmiss$ signal.}
\label{tab:ZW}
\end{table}

\subsection{Soft Muon Signals}

Similarly to the case of gluino production, we can also consider the additional features
of isolated soft muons from the decays of nearly degenerate gauginos. 
Due to the lack of boost effects, the result mostly depends on the mass splitting
between $\chi^\pm_1$/$\chi^0_2$ and $\chi^0_1$ states.
As considered in the last section, we intend to explore the signal with an isolated
muon in the hope to separate out the nearly degenerate gaugino production. 
The relevant leptonic decays of the  chargino and neutralino 
through the off-shell $W/Z$ yield typical  branching fractions as 
\be
\text{BR}(\chi^\pm_1\to \chi^0_1 \mu^\pm \nu_\mu)\simeq 11.1\%,~~ 
\text{BR}(\chi^0_2\to \chi^0_1\mu^+\mu^-)\simeq 3.3\% .
\ee
With these, we estimate that the $\chi^\pm_1\chi^0_2$ signal cross section is 
about 30 fb.  The signal can be roughly doubled if we also count for other 
channels of gaugino production.

It turns out that there are still substantial SM backgrounds with mono-jet$+\etmiss+\mu^{\pm}$
as that of Eq.~(\ref{eq:W}). We impose the selection cut as in Eq.~(\ref{eq:ETmcut}), and 
require that there be a soft muon satisfying the criterion described in Eq.~(\ref{eq:ptl}).
The entries in the last column in \ref{tab:ZW} compare these SM backgrounds as listed,
and the total background rate is about  800 fb.

To compare with the situation in the last section, we estimate that with 
an integrated luminosity of 10 fb$^{-1}$,  we can reach about $7\sigma$ sensitivity 
for $M_1=120$ GeV,  while  $S/B\sim 4\%$. 
Despite the improvement with the soft muon requirement, 
one would have to keep the systematic effects well under control to claim a discovery.

\section{Gaugino Pair Production via WBF: Two jets + $\etmiss$}

Given the difficulty for the observation of the signal from mono-jet plus $\etmiss$, we next
consider gaugino pair production  
from weak gauge  boson fusion (WBF). The rather distinctive
jet kinematics may provide sufficient discrimination power to extract 
the signal. WBF gaugino pair production at the LHC has been studied for pure wino
LSP case \cite{Datta:2001hv} and for general SPS points \cite{Cho:2006sx}. 
In our degenerate gaugino cases, charginos in WBF production will not get highly
boosted so the leptons are mostly  soft. Therefore, we will
focus on two very energetic forward/backward jets $+\cancel{E}_T$ final state which is
similar to the invisible Higgs search \cite{Eboli:2000ze} but at much smaller rates.
Similarly to what we have proposed in the gluino or mono-jet case, one can also search for
soft leptons in the 2 jets+$\cancel{E}_T$ samples.  

\begin{figure}[tb]
\includegraphics[scale=0.3,width=8cm]{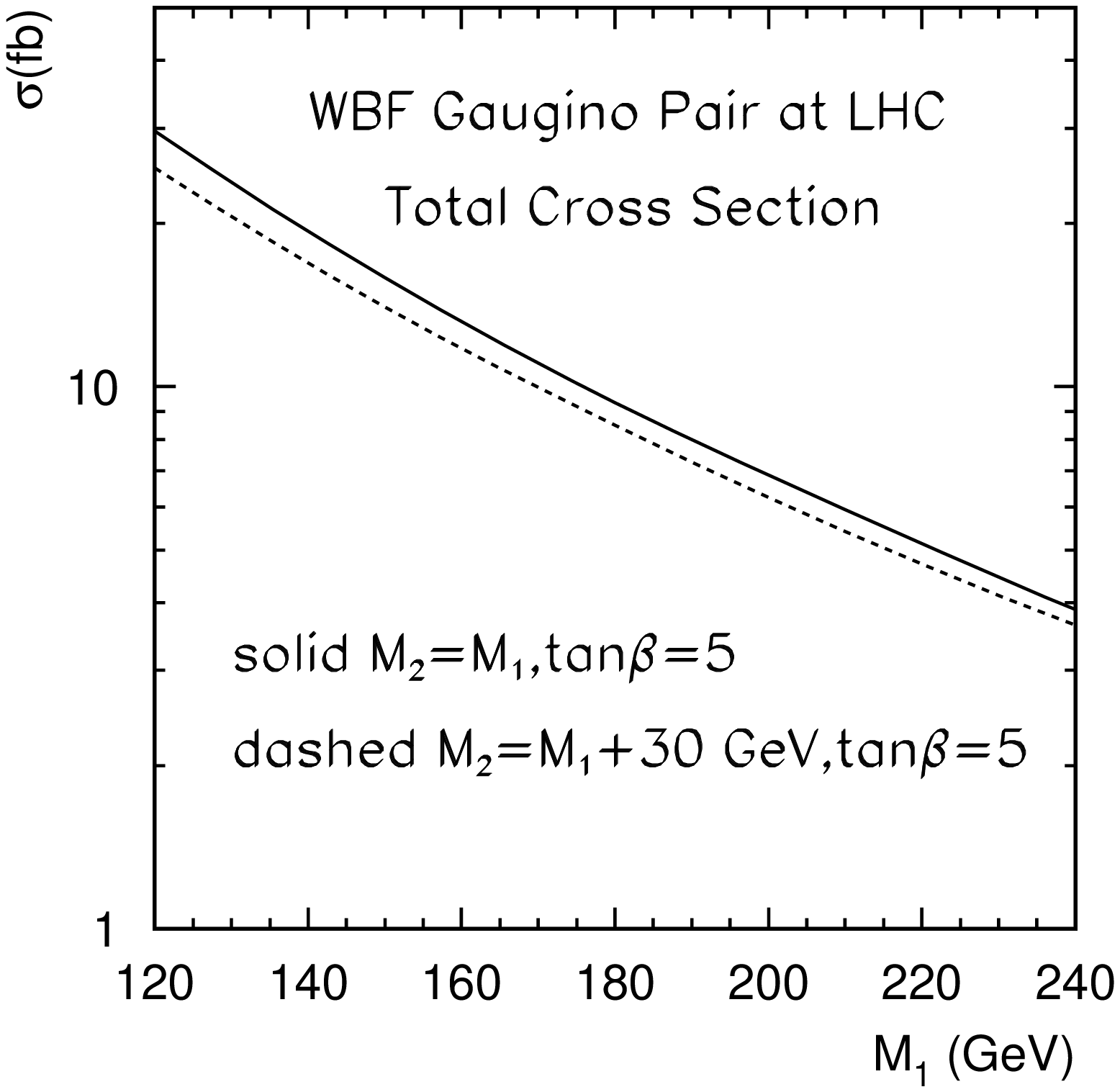}
\includegraphics[scale=0.3,width=8cm]{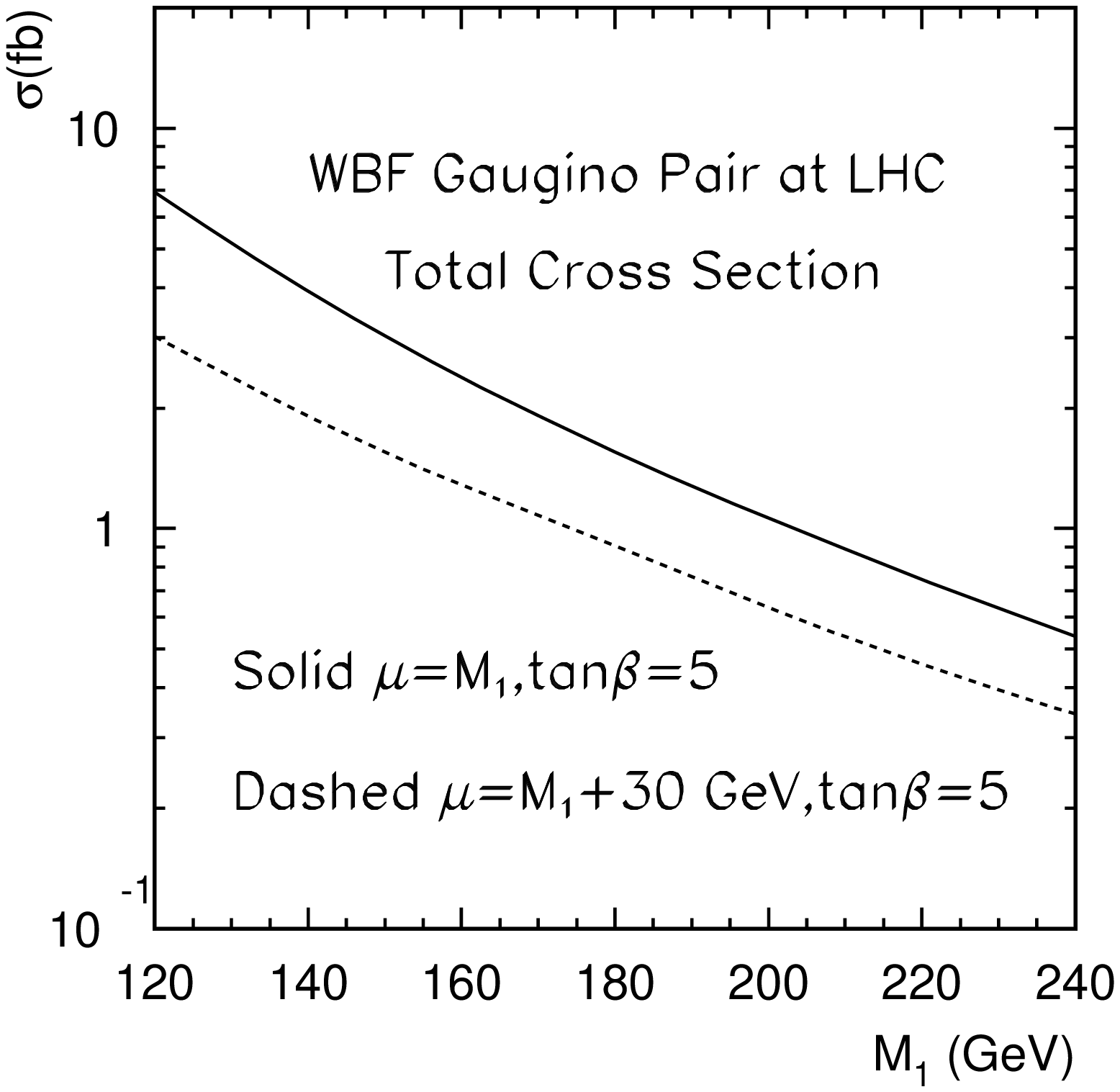}
\caption{
Total cross sections for the WBF signal with the jet-tagging cuts in Eq.~(\ref{eq:tagWBF})
versus $M_1$ for (a)  $M_2=M_1,\ M_1+30$ GeV and $\mu\to \infty$,
and (b) $\mu=M_1,\ M_1+30$ GeV and $M_2 \to \infty$. 
The leading channels of gaugino pair production are all summed over.}
\label{fig:m2muWBF}
\end{figure}

In addition to be an important discovery channel, the observation of
the WBF process also helps to reveal the identies of the lower lying
gaugino states. For example, the pure bino LSP will have a
vanishing WBF production rate. In principle, one can also distinguish the mixed bino-wino and the mixed bino-higgsino 
cases since they predict different production rates.
Similar to the other production channels, identification of soft leptons will both add a useful discovery channel, and provide 
crucial information of the gaugino spectrum. 


\subsection{2 Jets +  $\etmiss$ in WBF} 

\begin{figure}[tb]
\includegraphics[scale=1,width=8cm]{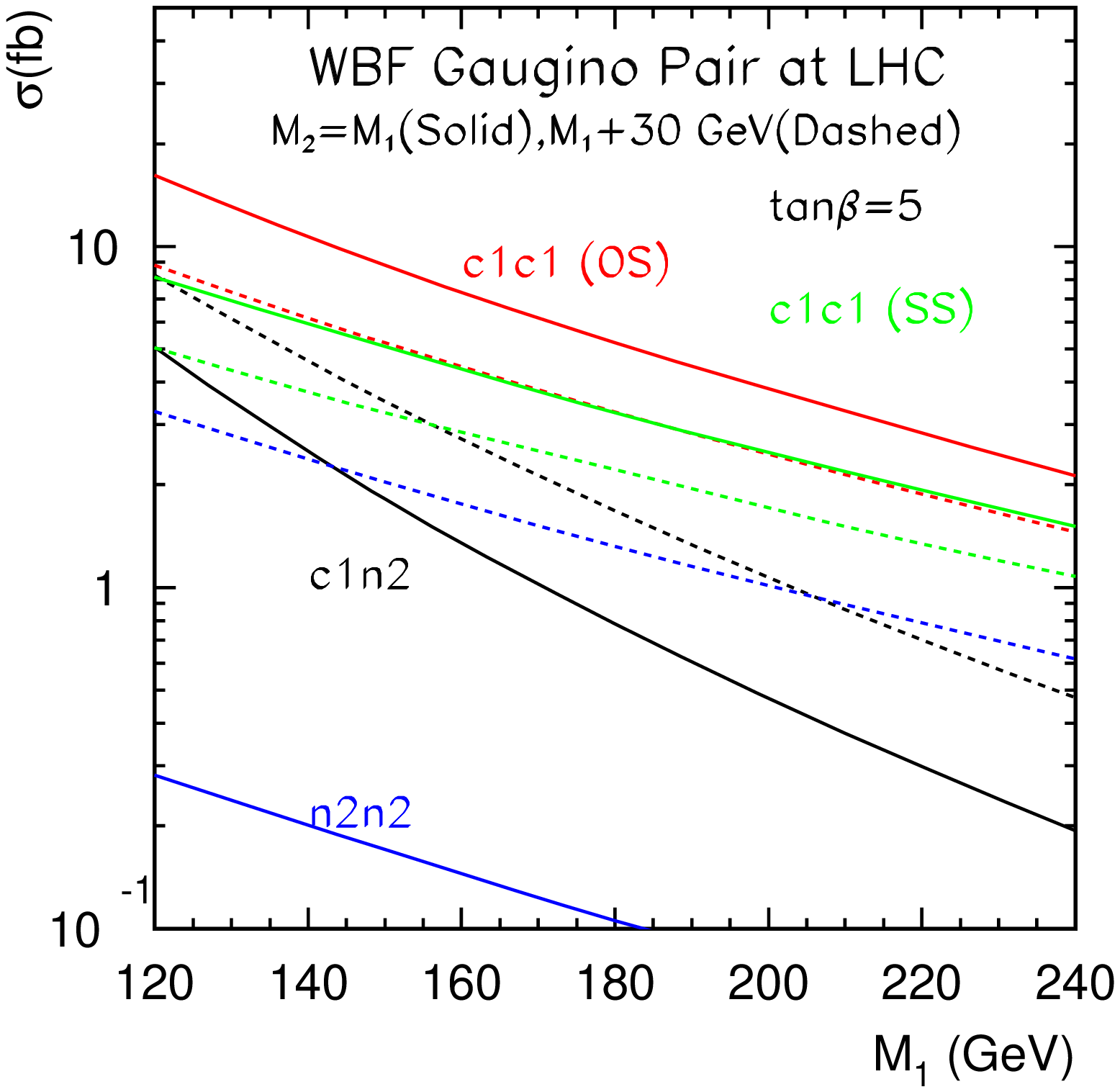}
\includegraphics[scale=1,width=8cm]{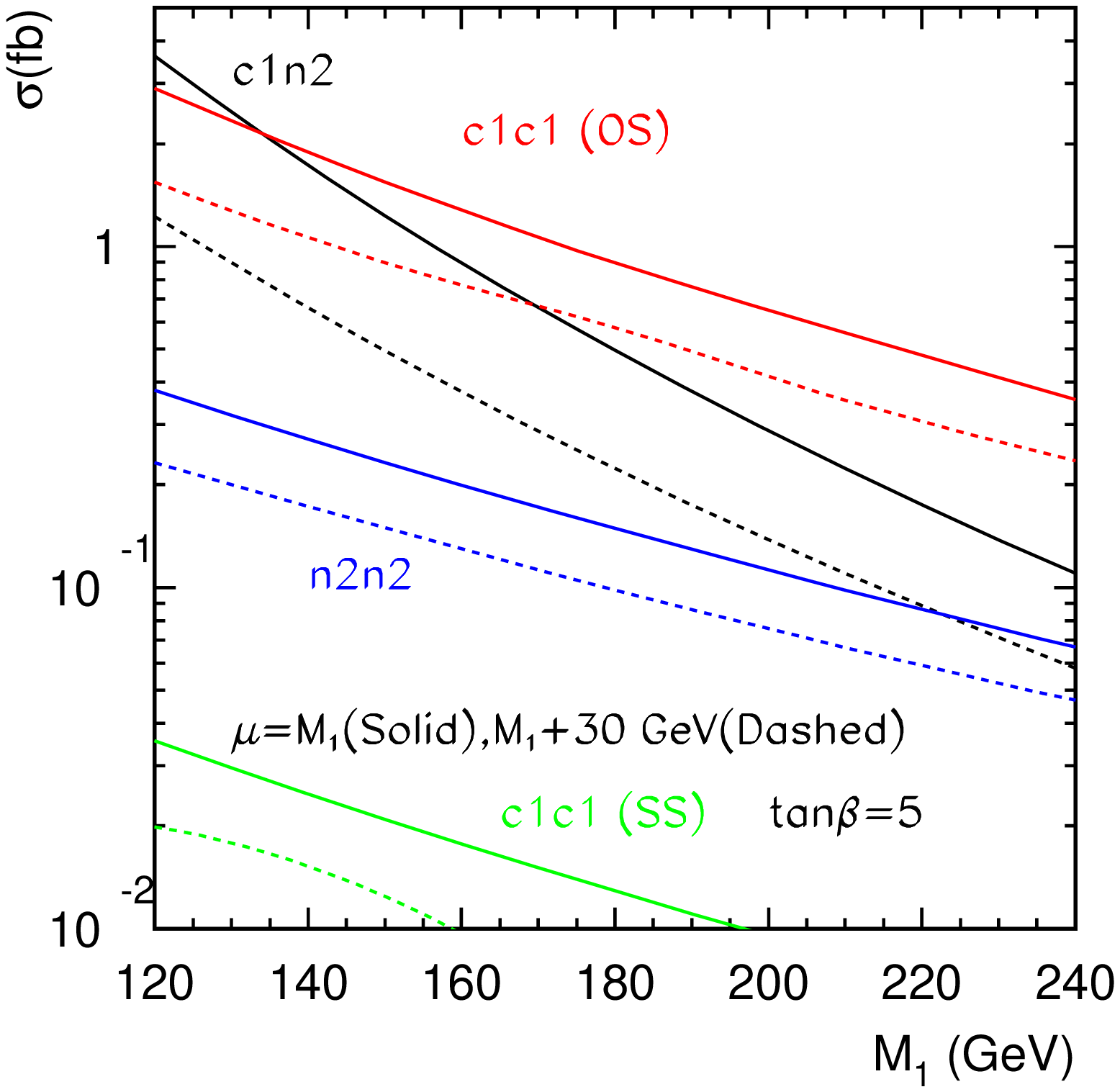}
\caption{
Total cross sections for the WBF signal with the  tagging cut in Eq.~(\ref{eq:tagWBF}),
for (a) $\tilde B-\tilde W$ mixing with $M_2=M_1$ (solid line) and $M_2=M_1+30$~GeV (dashed line),  
and (b) $\tilde B-\tilde H$ mixing  with $\mu=M_1$ (solid line) and $\mu=M_1+30$~GeV (dashed line).
Labels in the figure denote different productions channels  $jj\chi^+_1\chi^-_1$: c1c1 (OS); 
$jj\chi^\pm_1\chi^\pm_1$: c1c1 (SS); $jj\chi^\pm_1\chi^0_2$: c1n2; $jj\chi^0_2\chi^0_2$: n2n2.
 } 
\label{wbfrate}
\end{figure}

Because of the characteristic features of the WBF kinematics \cite{centralveto}, 
we demand the basic cuts for the two tagged forward-backward jets
\be
E_T^j > 30\ {\rm GeV},\quad  |\eta_j| < 5.0,\quad \Delta R_{jj} >0.7.
\label{eq:tagWBF}
\ee
The signal rates including all the gaugino pairs in the final states are shown by the solid curves in 
Fig.~\ref{fig:m2muWBF} for (a) $\tilde B-\tilde W$ mixing, and (b)  $\tilde B-\tilde H$ mixing. 
We see that the signal cross sections in the parameter region of our interest are of the order of 
$4-30$ fb for the case of $\tilde B-\tilde W$ mixing, and $0.5-7$ fb for $\tilde B-\tilde H$ mixing. 
The rate is typically smaller than that of gaugino pair plus a mono-jet signal studied in the last
section by $1-2$ orders of magnitude.
The separate inidvidual channels are shown by the solid curves in Fig.~\ref{wbfrate},
again for (a) $\tilde B-\tilde W$ mixing and (b)  $\tilde B-\tilde H$ mixing. 
A light  $\tilde W^{\pm}$ scenario from $\tilde B-\tilde W$ mixing is significantly larger than
the light $\tilde H^{\pm}$ scenario from  $\tilde B-\tilde H$ mixing. 
The opposite-sign (OS) pair production of $\chi^{+}\chi^{-}$ is always a leading channel.
The same-sign (SS) pair production of $\chi^{\pm}\chi^{\pm}$, however, is only large for a light $\tilde W^{\pm}$,
but highly suppressed for a light $\tilde H^{\pm}$. 

\begin{figure}[tb]
\includegraphics[scale=1,width=9.5cm]{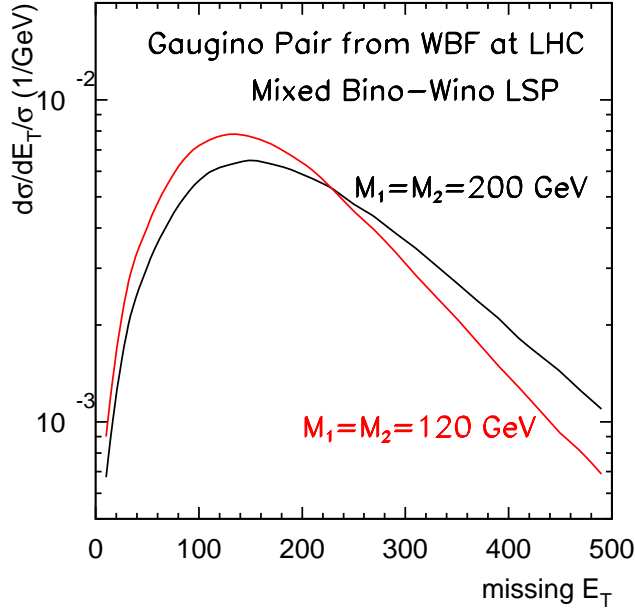}
\caption{Normalized $\cancel{E}_T$ distribution for WBF with
 $M_1=M_2=120$ GeV (red) and  $M_1=M_2=200$ GeV (black). }
 \label{fig:etmWBF}
\end{figure}
For the signal, significant $\etmiss$ arises from the missing gaugino pairs. As seen in Fig.~\ref{fig:etmWBF}
with two representative scales 120 and 200 GeV, the heavier
gauginos lead to somewhat harder $\etmiss$ spectrum. 
In the case
of near degeneracy under consideration, additional  cascades involving different gaugino states have negligible effect on the 
$\etmiss$ spectrum.   By applying a large $\etmiss$ cut, 
one can dramatically reduce the SM background.  For this as 
well as for a triggering purpose, we demand large missing transverse energy
\be
\etmiss > 100\ {\gev}.
\label{eq:etmissWBF}
\ee
\begin{figure}[tb]
\includegraphics[scale=1,width=7cm]{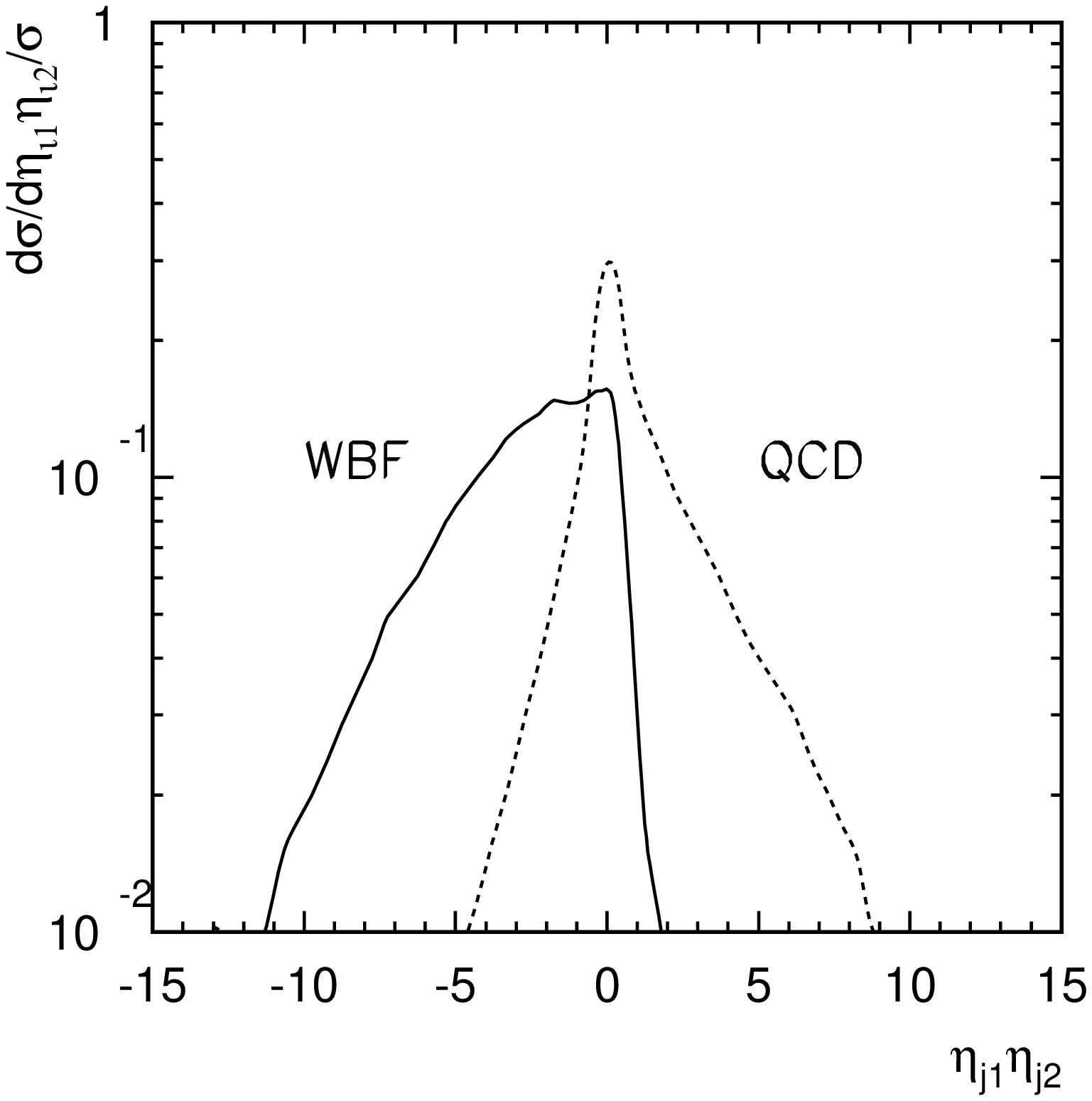}
\includegraphics[scale=1,width=7cm]{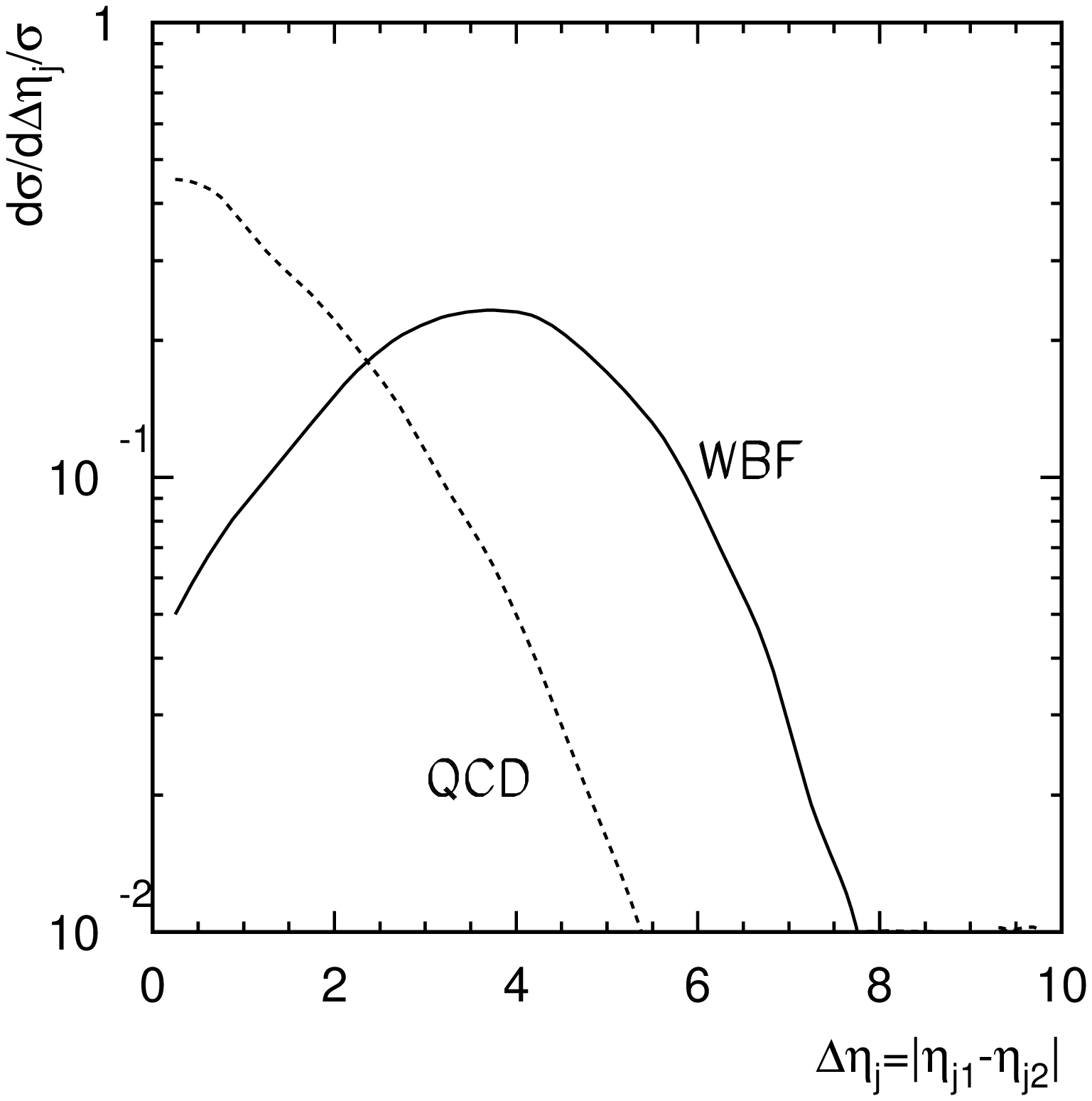}
\includegraphics[scale=1,width=7cm]{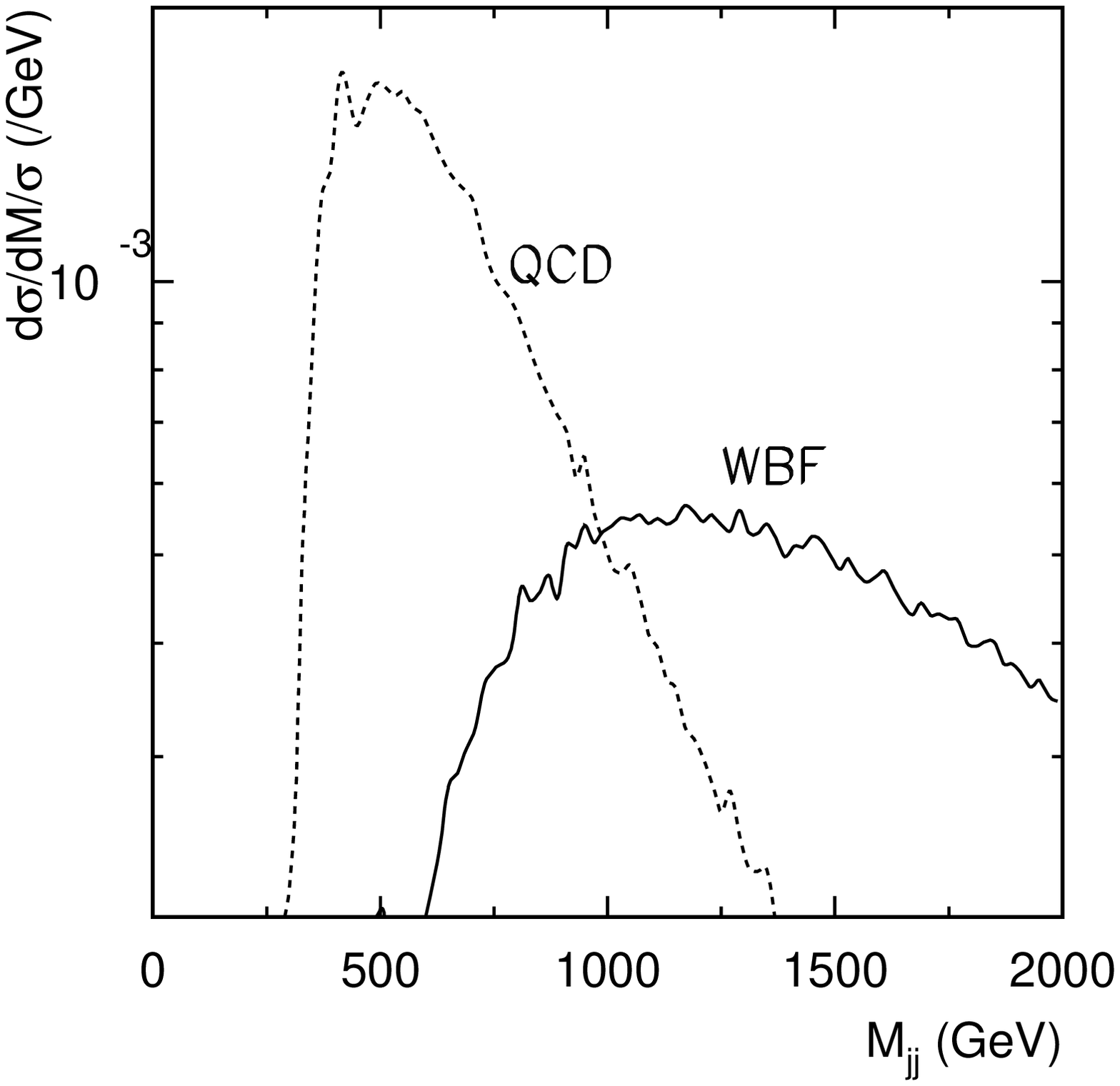}
\caption{Normalized distributions for WBF production of $jj\chi^+\chi^-$ (solid) 
and the QCD background $jjW$ (dashed) with basic cuts applied,
(a) $\eta_{j_1} \eta_{j_2}$, (b) $|\eta_{j_1} -\eta_{j_2}|$ and (c) the di-jet 
 invariant mass $M_{JJ}$. } 
 \label{fig:etaWBF}
\end{figure}
\noindent
To further illustrate the striking feature of the WBF kinematics, we look into the two pseudo-rapidities of
the two tagged jets $\eta_{j_1},\ \eta_{j_2}$. The two jets are typically in the opposite hemispheres with
respect to the central region $\eta=0$. 
Figures \ref{fig:etaWBF}(a)$-$(b) show the normalized distributions of 
$\eta_{j_1}\cdot \eta_{j_2}$ and $|\eta_{j_1} -\eta_{j_2}|$ for the WBF signal, 
compared with the leading QCD background $jjW$. 
We thus impose the additional cuts on them
\be
\eta_{j_1} \eta_{j_2} <0,\quad  |\eta_{j_1}- \eta_{j_2}|>4.4.
\ee
%
%
\noindent
The large rapidity separation of the forward-backward jets implies a larger
invariant mass of the di-jet system, in comparison with the QCD
background, as shown in Fig.~\ref{fig:etaWBF}(c). 
We thus impose an additional cut on the di-jet mass, 
\begin{equation}
M_{JJ} > 1200 ~{\rm GeV}.
\end{equation}
We find that tightening up the jet $p_T$ could further improve the signal-to-background ratio,
and we thus include one more cut 
\be
p_{T}^{J} > 60\ {\rm GeV}.
\ee
in our background estimates and our final analysis of the reach. 
%

\begin{table}[tb]
\begin{tabular}{| c || c  | c | c| c || c||}
    \hline
     Processes (fb) & Basic Cuts   & $\eta_{j1}\eta_{j2}<0$   & $M_{JJ}$  & $p_T^J$   \\
    &Eqs.~(\ref{eq:tagWBF}), ~(\ref{eq:etmissWBF}) & $|\eta_{j1}-\eta_{j2}|>4.4$ & $> 1200$~GeV & $>60$~GeV \\
    \hline
       $Zjj$ (EW)& 1400 & 170  & 120  & 87  \\
       $P_{\rm surv} \sigma$ & 1200 & 140  & 97 & 71  \\

   \hline 
       $Zjj$ (QCD) & 125$\times 10^{3}$ & 3100  & 970 & 520 \\
       $P_{\rm surv} \sigma$ & 35$\times 10^{3}$& 880 &  270 & 150 \\
  
  \hline

       $Zjj$ Total & 36$\times 10^{3}$ 
       & 1000 & 370 & 220  \\
  \hline 
       $Wjj$ (EW)& 200 & 38 & 27 & 20\\
       $P_{\rm surv} \sigma$ & 160 & 31  & 22 & 16 \\

   \hline 
       $Wjj$ (QCD)& 21$\times 10^{3}$& 630 & 230 & 120 \\
       $P_{\rm surv} \sigma$ & 6.0$\times 10^{3}$ &180 & 64 & 34 \\
  
  \hline

       $Wjj$ Total & 6.2$\times 10^{3}$ & 210  & 86 & 50 \\
  \hline 
  \hline

       Total BG& 42$\times 10^{3}$  & 1200 & 450 & 270 \\
       \hline
  \end{tabular}
  \caption{The SM background rates (in fb) of  two-jets plus $\cancel{E}_T$ with the consecutive acceptance cuts. 
  The rows indicated by $P_{\rm surv} \sigma$ denote the estimates after the central jet veto. } 
  \label{tab:nolepWBF}
\end{table}

The leading backgrounds are 
\begin{itemize} 
\item 2 Jets + $Z$ with $Z\rightarrow \nu\bar{\nu}$, both from QCD and from EW;
\item 2 Jets + $W^{\pm}$ with $W^{\pm}\rightarrow \nu X$, both from QCD and from EW; similar to the 
discussion in previos sections.
\end{itemize} 
Besides the kinematical cuts discussed above, we require that
there be no leptons within 
\be
p_T >10~ {\rm GeV},\quad  |\eta|< 3.0.
\ee
The QCD background, for which there is color exchange through the $t$-channel gluon, has more jet activity in the central region. 
The effect of a central jet veto has been analyzed  for various processes in Ref.~\cite{Eboli:2000ze}. 
From those analyses we infer veto survival probabilities  of $28\%$ for QCD $Zjj$ and $Wjj$, 
and $82\%$ for EW $Zjj$ $Wjj$
and $\chi_i\chi_j jj$. 
We summarize the acceptance of the  backgrounds with the consecutive cuts
in Table \ref{tab:nolepWBF}. 

\begin{figure}[tb]
\includegraphics[scale=1,width=8cm]{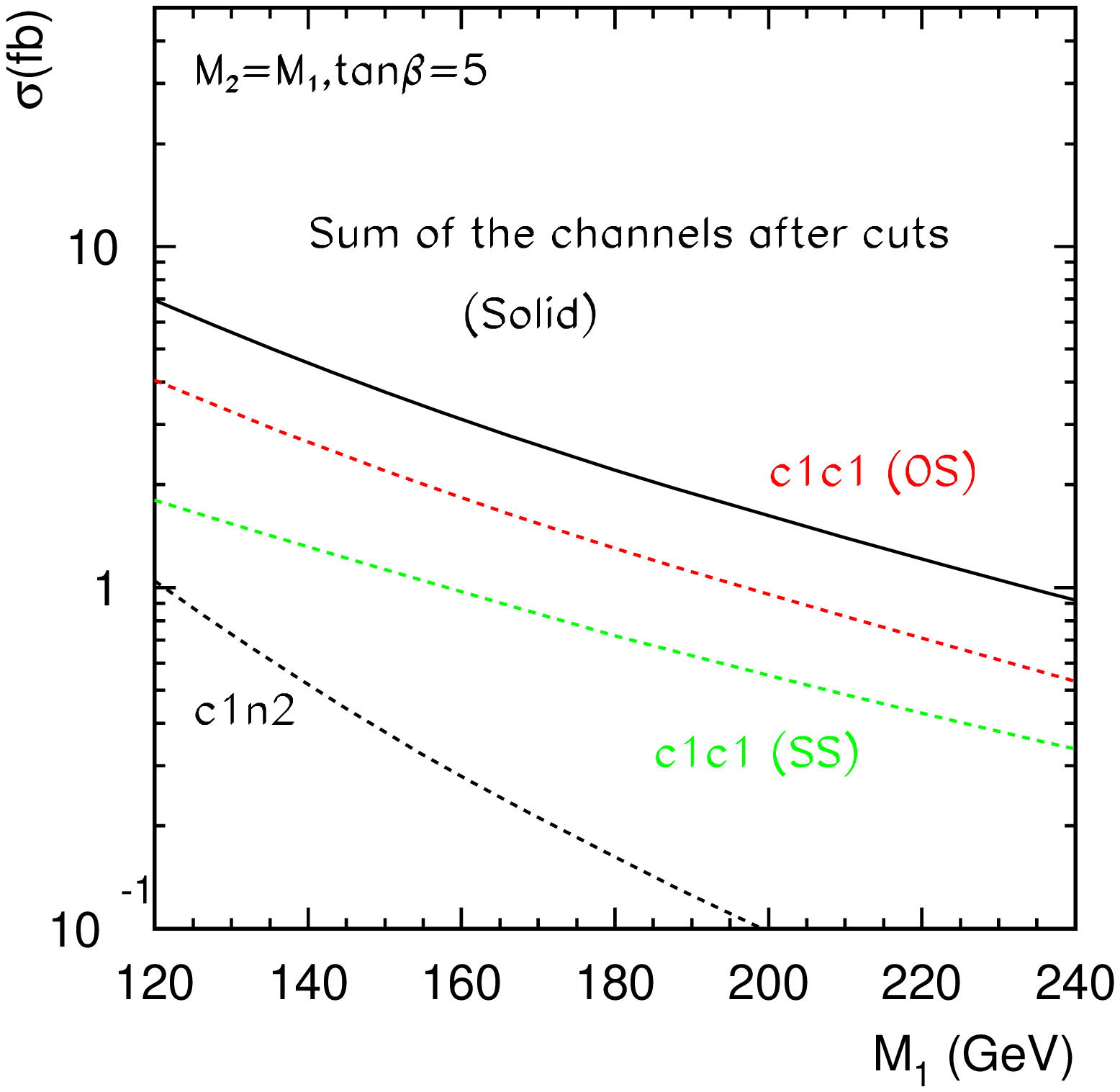}
\includegraphics[scale=1,width=8cm]{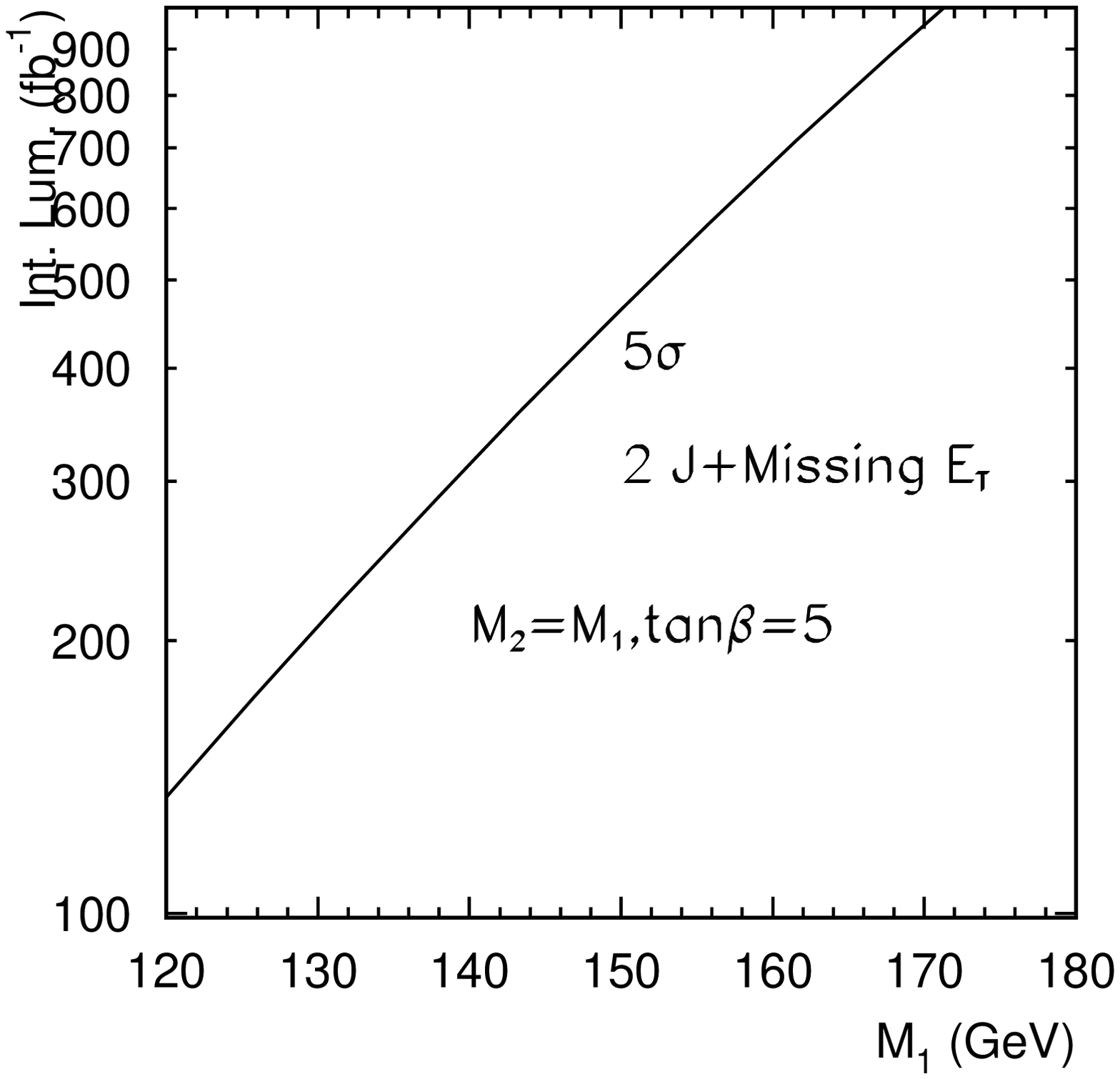}
\caption{
(a) Total cross sections for the WBF signal;
(b) Integrated luminosity needed to reach 5$\sigma$ $S/\sqrt{B}$
for $\tilde B-\tilde W$ mixing $M_1=M_2$.  For both (a) and (b), all the cuts used in Table~\ref{tab:nolepWBF} have been imposed. The 
labels in the figure denote different production channels: $jj\chi^+_1\chi^-_1$: c1c1 (OS); $jj\chi^\pm_1\chi^\pm_1$: c1c1 (SS); $jj\chi^\pm_1\chi^0_2$: c1n2; $jj\chi^0_2\chi^0_2$: n2n2.
 } 
\label{wbfratecut}
\end{figure}

To evaluate the signal observability, 
we study the signal after applying all the cuts described above for $M_1=M_2$, $\tan\beta=5$, $M_{\tilde{f}}=5$~TeV, and $\mu=1$~TeV. 
The resulting signal rates are shown in Fig.~\ref{wbfratecut} for the individual channels as well as the total sum (solid).  
Considering the backgrounds given in Table \ref{tab:nolepWBF}(a), we obtain the integrated luminosity needed to reach a $5\sigma$
statistical significance of the signal in Fig.~\ref{wbfratecut}(b). 
We see that, not surprisingly, that the signal observation is very challenging. For instance, 
the degenerate gaugino signals from the WBF for $\mlsp \simeq 145 $ GeV may be reached 
at $5 \sigma$ level with a high luminosity of $300$ fb$^{-1}$. But one must control the systematics very well
since $S/B\sim 2-3\%$ only.
Further refinement and improvement are possible such as exploiting the leptons in the events. We leave those
to a more comprehensive detector simulations. Instead, we only try to examine the events with soft muons next.

\subsection{Soft Muons}
As we described in the previous sections, for a specific window of mass splitting $\dm= m_{\chi^+_1}-m_{\chi^0_1}$, 
the search for isolated soft muons becomes an important handle to identify the nature of the neutralino and chargino states. 
 

Considering the WBF signal for the degenerate gauginos, the 
leading background with an isolated soft muon comes from  the process
2 jets + $W^{\pm}$ with $W^{\pm}\rightarrow \nu \mu$, with the soft muon satisfying Eq.~(\ref{eq:ptl}),
both from QCD jets and from EW quark scattering. 
The rates of these two SM background processes are summarized in Table~\ref{tab:lepWBF}. 
With respect to the previous section, the background situation is significantly improved. First,
here is no significant contribution from  $Z$ production due to the requirement both for large $\etmiss$ 
and a muon in the final state. Second, the background rates for the $Wjj$ production is reduced by
about a factor of four by the soft muon requirement. 

\begin{table}[tb]
\begin{tabular}{| c || c  | c || c | c || c||}
    \hline
     & $Wjj$ (EW) & $P_{\rm surv} \sigma$  & $Wjj$ (QCD) & $P_{\rm surv} \sigma$ & Total \\
    \hline
      $\sigma$ (fb)    & 4.5 & 3.7 & 36 &10 & 14 \\
       \hline
  \end{tabular}
  \caption{The SM backgrounds of two-jets$+ \cancel{E}_T$+ 1 soft muon after all cuts and the
  soft muon selection. 
The entries indicated by $P_{\rm surv} \sigma$ denote the estimates after the central jet veto.}  
  \label{tab:lepWBF}
\end{table}

Given that the branching fraction of $\chi^\pm_1\ (\chi^0_2)$ to a muon final state is $11\%\ (3\%)$, 
we can estimate the discovery potential for the isolated soft muon signal. 
If $M_2$ is around 120 GeV and the $\dm$ is sufficiently large,
one can expect the signal rate of about $2\times$BR$\times$(7 fb)$\approx 1.5$ fb. 
For 300 fb$^{-1}$ integrated
luminosity, we obtain $S/\sqrt{B} \sim 5\sigma$, while reaching $S/B\sim 10\%$. 
Although the statistical significance remains roughly the same before and after the soft muon
requirement, the the systematics as reflected in S/B are clearly improved.

\begin{figure}[tb]
\includegraphics[scale=1,width=9cm]{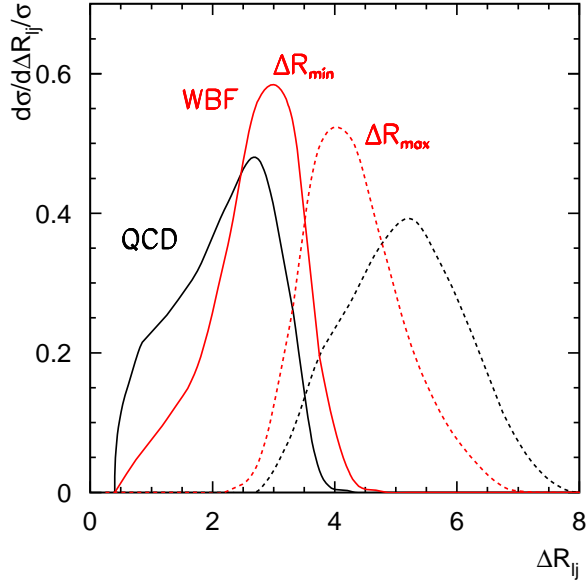}
\caption{Normalized $\Delta R$ distributions for a soft  muon with respect to the two jets, dashed line
for $\Delta R^{\rm max}_{\ell j}$ and solid line for $\Delta R^{\rm min}_{\ell j}$. 
The WBF signal (red curves) and QCD background
(black curves) are labeled respectively.
}
\label{fig:dr}
\end{figure}

Finally, we would like to point out that there are still some kinematic features that may be exploited
to further purify the isolated soft muon$+2j+\cancel{E}_T$ signal. 
In the QCD processes $Wjj/Zjj$, the $W/Z$ are radiated from a
quark line and therefore the lepton from the gauge boson decay is emitted close to a jet.
Indeed the $\Delta R_{\mu J_i}$ distribution of the QCD background peaks at small values of $\Delta R$, 
as shown in Fig.~\ref{fig:dr}.
One could consider to design a further cut such as  $\Delta R^{\rm min}_{\ell j}>2.0.$
%

Since the VBF channel has a very small production rate already at $E_{CM}=14 $ TeV, 
we expect that running at a lower $E_{CM}$ will render this channel unreachable.  
Since the effectiveness of this channel crucially relies on having high statistics with 
$\mathcal{O}(100$ fb$^{-1})$, we expect our numerical  study with higher 
$E_{CM}$ to be the most relevant one. Future luminosity upgrade of the LHC can 
certainly enhance the prospect of extracting important information from this channel. 

\section{Summary and Conclusions}

In this paper, we have considered the strategies for discovering 
electroweak gaugino states with nearly degenerate mass at the LHC. 
Significant efforts were made to explore the connection between 
the signal kinematics and the relevant masses for the gluino and gauginos,
hoping to probe the mass scales of the SUSY breaking and the dark matter. 
More specifically, we have focused on a scenario in which the mass splittings between the gauginos are in the range between a few GeV and roughly $30$ GeV. This situation is fairly generic in supersymmetric models that account for the correct density of dark matter. This is because weak-scale Higgsinos and winos annihilate very efficiently in the early universe, leading to an exceedingly small thermal relic density, while binos have the opposite problem and their typical relic density is too large. A certain degree of mass degeneracy is a way of solving this problem and of obtaining a viable supersymmetric dark-matter candidate.  
Motivated by this dark-matter connection,
we have considered scenarios in which either bino and Higgsino, or bino and wino, are nearly degenerate in mass. 
For concreteness, we studied  the later case in details for our presentation.  

In contrast to other well-studied nearly degenerate examples with more distinct collider signatures, such as the wino LSP scenario, the decay of heavier gauginos  in our case is prompt in the collider experimental environment. 
Therefore, we cannot rely on displaced vertices or long-lived charged tracks for signal identification. In this paper, we carried out comprehensive studies of three possible discovery channels for nearly degenerate gauginos. 
We demonstrated important kinematical features of the events in the hope to explore the relevant mass
scales such as the gluino mass and gaugino masses. We designed the optimal judicial cuts
and estimated the sensitivity reaches with respect to the SM background expectations.

\vskip 0.4mm
\noindent
{\bf (1) Production of gluino pair:} 

\noindent
The gauginos are produced as the decay products of the gluino. Due to the lack of hard leptons, the jets $+ \etmiss$ is probably the most useful channel in this case. 
We have demonstrated the dependence on the mass splittings in two benchmark cases of the gluino 
mass $M_3 = 500$ GeV  and 1 TeV. 
We found that, at $E_{CM} = 14 $ TeV, the reach of a $5\sigma$ discovery for the above mass benchmarks 
with $M_3 - M_{LSP} \geq 100$ GeV may required a luminosity of 50 pb$^{-1}$ and 50 fb$^{-1}$, respectively.
Running at a lower energy mainly affects the gluino production rate. 
%
We have also considered the case of looking one additional soft muon, or two same-sign muons, resulting from the decay of chargino or heavier second neutralino. We found that both leptonic channels can be useful in improving the signal-to-background ratio. Moreover, the presence of such soft leptons as part of the signal events provides a clear verification of the nearly degenerate gaugino scenario. 
The reach in this channel is mainly controlled by three factors: gluino production rate, $M_3 - M_{LSP}$, and gaugino mass splitting $\Delta M$. 
Assuming a signal being from supersymmetry, this channel would be sensitive to $M_3$ by measuring 
the production cross section plus the invariant mass spectrum,  and could offer an early opportunity of 
determining the $M_{LSP}$ by measuring $M_3 - M_{LSP}$.

\vskip 0.4mm
\noindent
{\bf (2) Production of gaugino pair plus jets:} 

\noindent
We considered the direct pair production of the gauginos. An additional hard jet is necessary to provide a trigger for this class of signal when the nearly degenerate gauginos may not result in easily detectable final state particles. 
This class of signal is perhaps the most model-independent search for dark matter candidates at colliders.
We found that the mono-jet $+ \etmiss$ signal is very challenging to search 
due to its rather small signal-to-background ratio and  kinematical similarity between
 the signal and the background. For instance, we can obtain a 
 $S/\sqrt B \sim 5 \sigma$ statistical significance with 10 fb$^{-1}$ in this channel for $M_{LSP} \simeq 120$ GeV,
 while the $S/B$ is only about 1$\%$. 
 Searching for additional soft muons in the events could significantly improve both statistical and systematic 
effects, reaching $S/B \sim 4\%$,  at some cost of the signal rate. 
Measuring $M_{LSP}$ in this channel requires a precise prediction of the jet energy spectrum for both signal and the background.
The production rate in this mono-jet channel falls very fast with increasing $M_{LSP}$. 
The discovery reach seems to be limited to about $M_{LSP} \sim 200$ GeV.

\vskip 0.4mm
\noindent
{\bf (3) Production of gaugino pair via weak boson fusion:} 

\noindent
We found producing gaugino pairs via weak boson fusion  
to be a very useful mechanism at low gaugino masses about $120$ GeV. 
We argue that these channels can be extremely  informative in probing the nature of the gaugino states.  In particular, these processes represent the inverse of the dominant gaugino annihilation in the early universe and thus contain some crucial information that can be used, in certain cases, to reconstruct the thermal relic density of gaugino dark matter. 
We found that the signals of large missing energy plus two forward-backward
 jets may be observable at a $4$--$6\sigma$ level above the large SM backgrounds 
 with an integrated luminosity of $100$--$300$ fb$^{-1}$. 
Demanding additional soft muons in the events could again improve both statistical and systematic 
effects, reaching about $S/B \sim 10\%$. 
Similar to the mono-jet signal, the signal rate for the VBF channels also drop rather fast with increasing $M_{LSP}$. 
We estimated the discovery reach to be once again about $M_{LSP} \sim 200$ GeV.

Given the strong motivation in considering the nearly degenerate gaugino scenario, we hope our study to be the first step in  dedicated efforts in discovering and understanding the rich signals in the variety of channels laid out in this paper. Although we have considered nearly degenerated bino-wino as our benchmark, we expect the lesson drawn from our study is applicable in the nearly degenerate bino-Higgsino case, as well as other scenarios where the mass splittings between electroweak-inos are on the order of GeV to 10s GeV. 


\acknowledgments
We thank Nima Arkani-Hamed, Howie Baer and Xerxes Tata for useful conversations.
The work of T.H.~was supported by the U.S.~Department of Energy
under grant DE-FG02-95ER40896, 
and that of L.-T. W.~was supported by the National
Science Foundation under grant PHY-0756966 and the Department of
Energy under grant DE-FG02-90ER40542. K.W.~was supported by the World Premier International
Research Center Initiative (WPI Initiative), MEXT, Japan.

\end{document}